\documentclass[fleqn,usenatbib]{mnras}
\usepackage{newtxtext,newtxmath}
\usepackage[T1]{fontenc}
\usepackage{ae,aecompl}
\usepackage{graphicx}	
\usepackage{amsmath}
\usepackage{times}
\usepackage{verbatim}
\usepackage{subfigure}
\usepackage{float}
\usepackage{color}

\usepackage{amssymb}

\title[Hot and counter-rotating disk galaxies in TNG]{Hot and counter-rotating star-forming disk galaxies in IllustrisTNG and their real-world counterparts}
\author[S. Lu et al.]
{Shengdong Lu$^{1}$\thanks{E-mail: \url{lushengdong@tsinghua.edu.cn}},
Dandan Xu$^{1}$\thanks{E-mail: \url{dandanxu@tsinghua.edu.cn}},
Yunchong Wang$^{2}$,
Yanmei Chen$^{3}$,
Ling Zhu$^{4}$,
\and
Shude Mao$^{1}$,
Volker Springel$^{5}$,
Jing Wang$^{6}$,
Mark Vogelsberger$^{7}$,
Lars Hernquist$^{8}$
\\
\\
$^{1}$Department of Astronomy, Tsinghua University, Beijing 100084, China\\
$^{2}$Kavli Institute for Particle Astrophysics and Cosmology, Physics Department, Stanford University, Stanford, CA 94305 \\
$^{3}$Department of Astronomy, Nanjing University, Nanjing, China \\
$^{4}$Shanghai Astronomical Observatory, Chinese Academy of Sciences, 80 Nandan Road, Shanghai 200030, China\\
$^{5}$Max-Planck-Institut f\"ur Astrophysik, Karl-Schwarzschild-Str. 1, D-85748, Garching, Germany\\
$^{6}$Kavli Institute for Astronomy and Astrophysics, Peking University, Beijing 100871, China\\
$^{7}$Kavli Institute for Astrophysics and Space Research, Department of Physics, MIT, Cambridge, MA 02139, USA\\
$^{8}$Harvard-Smithsonian Center for Astrophysics, 60 Garden Street, Cambridge, MA 02138, USA\\
}

\date{Accepted ***. Received ***; in original form ***}

\pubyear{2020}

\begin{document}
\label{firstpage}
\pagerange{\pageref{firstpage}--\pageref{lastpage}}
\maketitle
\begin{abstract}
A key feature of a large population of low-mass, late-type disk galaxies are star-forming disks with exponential light distributions. They are typically also associated with thin and flat morphologies, blue colours, and dynamically cold stars moving along circular orbits within co-planar thin gas disks. However, the latter features do not necessarily always imply the former, in fact, a variety of different kinematic configurations do exist. In this work, we use the cosmological hydrodynamical IllustrisTNG Simulation to study the nature and origin of dynamically hot, sometimes even counter-rotating, star-forming disk galaxies in the lower stellar mass range (between $5\times 10^9\,\mathrm{M_{\odot}}$ and $2\times 10^{10}\,\mathrm{M_{\odot}}$). We find that being dynamically hot arises in most cases as an induced transient state, for example due to galaxy interactions and merger activities, rather than as an age-dependent evolutionary phase of star-forming disk galaxies. The dynamically hot but still actively star-forming disks show a common feature of hosting kinematically misaligned gas and stellar disks, and centrally concentrated on-going star formation. The former is often accompanied by disturbed gas morphologies, while the latter is reflected in low gas and stellar spins in comparison to their dynamically cold, normal disk counterparts. Interestingly, observed galaxies from MaNGA with kinematic misalignment between gas and stars show remarkably similar general properties as the IllustrisTNG galaxies, and therefore are plausible real-world counterparts.  In turn, this allows us to make predictions for the stellar orbits and gas properties of these misaligned galaxies.
\end{abstract}


\section{Introduction}
\label{sec:introduction}
Star-forming disk galaxies with exponential surface brightness
profiles are often associated with thin and flat morphologies, blue
colours, and dynamically cold stars moving along circular orbits within
the co-planar thin gas disks. This is a standard picture inherited
from the disk formation theory, which posits that a proto dark-matter
halo gains its initial angular momentum through interactions with its
surrounding gravitational tidal fields until
turnaround~\citep{Hoyle(1949),Peebles(1969),Doroshkevich(1970),Fall(1983),White(1984),Catelan_and_Theuns(1996)},
after which the subsequent virialization preserves angular momentum,
and the gas settles into a rotation-supported disk with size related
to spin~\citep{Fall_and_Efstathiou(1980),Mo_et_al.(1998)} as it
radiatively cools down and accretes towards the halo
center~\citep[e.g.][]{Keres_et_al.(2005),Dekel_et_al.(2006)},
Correspondingly, subsequent generations of stars that form inside the
rotating gaseous disk naturally inherit a tangential motion and in
this way build up a flat and dynamically cold disk component of the
galaxy (e.g. \citealt{Zavala_et_al.(2015)}).

Recent hydrodynamic simulations have allowed us to follow the detailed
evolution of galaxies during the non-linear growth era, and have shown
that various processes complicate the simple picture described above
(e.g. \citealt{Zavala_et_al.(2015),Zjupa_and_Springel(2017)} and
references therein): angular momenta can be lost and regained due to
galactic winds and accretion, can be transferred between dark matter,
gas and stars, can be redistributed from one part of the galactic halo
to the other, or be modified by merger processes. As a consequence,
galaxies can transform between different morphologies and kinematics
while varying their star-forming activities accordingly. It is not
entirely clear how a dynamically cold star-forming disk galaxy starts
getting heated up and becomes dynamically hot, eventually ceasing its
star-formation and acquiring a spheroidal/elliptical shape. This
process is likely to proceed along multiple paths under different
circumstances.

In~\citet{Xu_et_al.(2019)}, the luminosity/stellar contributions of
different dynamical components (in terms of orbit circularity) were
calculated for galaxies from one of the currently most advanced
cosmological hydrodynamic models -- the IllustrisTNG Simulation~\citep{Marinacci_et_al.(2018),Naiman_et_al.(2018),Nelson_et_al.(2018),Pillepich_et_al.(2018b),Springel_et_al.(2018)}, which was compared to a sample of 260 CALIFA
galaxies~\citep{Falcon-Barroso_et_al.(2017),Zhu_et_al.(2018)}, showing
general agreement in many properties. Interestingly, a fraction of the
simulated lower-mass, late-type galaxies (within a stellar mass range
between $5\times 10^9\,\mathrm{M_{\odot}}$ and
$2\times 10^{10}\,\mathrm{M_{\odot}}$) with active star-formation and
exponential light profiles (i.e., disk-dominated morphologies),
exhibited however fairly high (low) fractions of dynamically hot
(cold) stellar orbits. From the observational perspective, this has also been seen in star-forming galaxies within this mass range. For example, in \citet{Zhu_et_al.(2018)}, some of the lower-mass and star-forming CALIFA galaxies indeed have relatively low cold orbit fraction but higher hot orbital fraction -- a feature not fully compatible with
usual disk-dominated galaxies.

In this study, we investigate the nature and origin of these objects
further using the IllustrisTNG Simulation. We find that these systems
show a common feature of hosting kinematically misaligned gas and
stellar disks, as well as centrally-concentrated on-going star
formation. The former is often accompanied by disturbed gas
morphologies to a certain extent, whereas the latter is consistent
with low gas and stellar spins. In comparison, their dynamically cold
star-forming disk counterparts exhibit well-established and co-planar
gaseous disks and extended on-going star-forming activity.

Indeed, various forms of misalignments have been widely reported in
these systems, between gas and stars, as well as between morphology
and kinematics. An extreme case is when the spin of the gas disk lies
in completely the opposite direction to that of existing stars, which
is dynamically more stable than a misalignment at any other
angles~\citep{Hunter_and_Toomre(1969),Tohline_and_Durisen(1982)}. Under
such circumstances, the current in-situ star-formation naturally
results in the youngest stellar population inheriting the gas orbital
motion, and thus moving against the bulk spin dominated by the older
stellar population; the galaxy can thus also possess a good fraction
of counter-rotating stellar
orbits~\citep[e.g.][]{Algorry_et_al.(2014),Starkenburg_et_al.(2019)}. Such
a dynamically mixed configuration can be unveiled more specifically by conducting
an orbit-based modeling~\citep[e.g.][]{Zhu_et_al.(2018)}. Large misalignment between the kinematic axes of gas and stars or counter-rotation among different stellar components are also smoking gun signatures for such a system and can be easily recognized from the velocity maps of galaxies (e.g., \citealt{Coccato_et_al.(2013),Coccato_et_al.(2015),Katkov_et_al.(2016),Pizzella_et_al.(2018),Li_et_al.(2021)}).

Throughout the years, the origin of such misalignments has been suggested to
be related to galaxy mergers~\citep[e.g.][]{Balcells_and_Stanford(1990),Hernquist_and_Barnes(1991), Barnes_and_Hernquist(1996),Crocker_et_al.(2009)}, satellite accretion~\citep[e.g.][]{Thakar_and_Ryden(1996),Puerari_and_Pfenniger(2001), Algorry_et_al.(2014)}, or misaligned gas accretion (e.g. \citealt{Pizzella_et_al.(2004),Roskar_et_al.(2010),van_de_Voort_et_al.(2015),Graham_et_al.(2017),Starkenburg_et_al.(2019),Khoperskov_et_al.(2021)}), which are not
necessarily exclusive to one another.

In this study, we find evidence that the dynamically hot
configurations are often linked to recent galactic interactions and
mergers, which trigger a subsequent misalignment between the spins of gas
and stars in the inner regions of galaxies. Our study also reveals
that being dynamically hot is typically a transient state rather than
a long-term evolutionary phase for star-forming disk galaxies.
Interestingly, such an ``unfortunate'' disk galaxy is also able to get
``healthy'' again. Through our comparison, remarkable similarities in the
distributions of many observational properties are found between
misaligned star-forming galaxies from the MaNGA survey and the
dynamically hot star-forming disks from the IllustrisTNG Simulation.

The paper is organized as follows. In Section~\ref{sec:data}, we
introduce how we calculate relevant properties for galaxies from the
IllustrisTNG Simulation (Section.~\ref{sec:proptcal}), and how we
select sample galaxies for this study
(Section~\ref{sec:selection}). In Section~\ref{sec:look_like}, we
present how typical dynamically hot star-forming disks look like,
considering their stellar and gas components, their morphologies
(Section~\ref{sec:mor_of_gas_and_star}), kinematics
(Section~\ref{sec:kin_of_gas_and_star}), and star-formation activities
(Section~\ref{sec:SF_activity}).  In Section~\ref{sec:how_form}, we
discuss the formation and evolution of dynamically hot star-forming
disk galaxies in comparison to their normal disk counterparts. In
Section~\ref{sec:observation}, we present several key observational
properties of the real-world counterparts to the simulated galaxies
for comparison. In particular, predictions are made for the misaligned
galaxies with respect to stellar-orbit and gas observations. Finally,
conclusions and some further discussions are given in
Section~\ref{sec:conclusion}.

In this work, we adopt the cosmology used in the IllustrisTNG
simulation, which is based on results of the Planck
experiment~\citep{Planck_Collaboration(2016)}: flat universe geometry
is assumed with a total matter density of $\Omega_{\rm m} = 0.3089$
(with a baryonic density of $\Omega_{\rm b} = 0.0486$), a cosmological
constant of $\Omega_{\Lambda} = 0.6911$, and a Hubble constant
$h = H_0/(100\,{\rm km s}^{-1} {\rm Mpc^{-1}}) = 0.6774$.

\section{Methodology}
\label{sec:data}
\textit{The Next Generation Illustris Simulations} (IllustrisTNG, TNG
hereafter; \citealt{Marinacci_et_al.(2018),Naiman_et_al.(2018),Nelson_et_al.(2018),Nelson_et_al.(2019b),Pillepich_et_al.(2018b),Pillepich_et_al.(2019),Springel_et_al.(2018)}) are a suite of state-of-the-art magneto-hydrodynamic cosmological galaxy formation simulations carried out in large cosmological volumes with the moving-mesh code \textsc{arepo} \citep{Springel(2010)}.  In
this study, we use the full-physics version with a cubic box of $110.7\,\mathrm{Mpc}$ side length (TNG100), which has a mass resolution for baryonic and dark matter of $m_{\rm baryon}=1.4\times10^6\,{\rm M_{\odot}}$ and
$m_{\rm DM}=7.5\times10^6\,{\rm M_{\odot}}$, respectively. The gravitational softening length for dark matter and stellar particles is $\epsilon_{\rm softening} = 0.74\,\mathrm{kpc}$. Galaxies in their host dark matter halos are identified using the {\sc subfind}
algorithm \citep{Springel_et_al.(2001),Dolag_et_al.(2009)}. General galaxy properties have been calculated and publicly released by the TNG collaboration\footnote{\url{http://www.tng-project.org/data/}}
\citep{nelson_et_al.(2019a)}. In the following, we explain how several specific further galaxy properties are calculated, and how three key galaxy samples are composed for this study.

\subsection{Galaxy property calculation}
\label{sec:proptcal}
Star-formation rate, morphology and kinematics are three main aspects
describing the state of a given galaxy. Below we detail the definition
of a few further galaxy properties that we calculate using the
constituent particle/cell information:
\begin{enumerate}

\item specific star-formation rate $\mathrm{sSFR}$ in the past 1 Gyr,
  calculated within a 2D aperture of radius $2R_{\rm hsm}$, where
  $R_{\rm hsm}$ is the 3D half-stellar mass radius measured from the
  center of a given galaxy.

\item a bulge-to-total luminosity ratio $L_{\rm dev}/L_{\rm tot}$,
  given by the de Vaucouleurs~\citep{de_Vaucouleurs(1948)} to
  total luminosity ratio, which is obtained from fitting a two
  component model composed of a de Vaucouleurs and an exponential
  profile to the radial surface brightness distribution of the
  elliptical isophotes (see~\citealt{Xu_et_al.(2017)} for details).

\item the surface stellar mass density calculated within an aperture of radius $R_{\rm hsm}$, $\Sigma_{\ast,R_{\rm hsm}}$.

\item the shortest-to-longest axis ratios $(c/a)_{\ast}$ and $(c/a)_{\,\rm
  gas}$ of a galaxy's stellar and gaseous disks, respectively,
  obtained through finding eigenvectors of the associated inertia
  tensors~\citep{Allgood_et_al.(2006)}, defined using particles/cells within a
  chosen 3D radius from the galaxy center. This radius is set to be
  the smaller one of $3R_{\rm hsm}$ and 30 kpc, in order to
  restrict the calculation to the visible galaxy region.

\item the stellar and HI gas (here the HI gas refers to neutral hydrogen, without distinguishing molecular hydrogen and atomic hydrogen) line-of-sight velocity maps within a square of $[-3R_{\rm hsm},\,3R_{\rm hsm}]^2$ from the galaxy center in a given projection, weighted by stellar luminosity and HI mass, respectively.

\item spin parameters: specific angular momenta for both stellar
  ($j_{\ast}$) and gaseous ($j_{\rm gas}$) components calculated
  within $3R_{\rm hsm}$.

\item luminosity fractions of stars in ``cold'', ``hot'' and
  ``counter-rotating cold'' orbits. An orbital type is defined using
  stellar instantaneous circularities
  $\lambda_z \equiv L_{\rm z} / J_{\rm c}$, where $L_{\rm z}$ is a stellar
  particle's angular momentum component in the direction of the
  galaxy's shortest principal axis $z$ found for the stellar
  component; and $J_{\rm c}$ is the maximum $L_{\rm z}$ (corresponding to
  the circular orbit) among all stars that have the same binding
  energy as the given stellar particle. The three fractions are
  denoted as $f_{\rm cold}$, $f_{\rm hot}$, and $f_{\rm ccd}$,
  respectively, all evaluated within a 3D radius of $2R_{\rm hsm}$
  from the galaxy center, defined according to the following criteria
  (for more details, see~\citealt{Xu_et_al.(2019)}):

\begin{enumerate}
\item cold component: $\lambda_z>0.8$;
\item hot component: $-0.25<\lambda_z<0.25$;
\item counter-rotating cold component: $\lambda_z<-0.6$\footnote{We
  note that if $-0.8$ is adopted instead of $-0.6$ as the criteria for
  the sake of symmetry, a significant fraction of counter-rotating
  galaxy sample (see Section~\ref{sec:selection}) will be gone, as
  galaxies simply do not show such extremely negative circularities in
  any significant fractions of their stellar populations.}.
\end{enumerate}
\end{enumerate}

We note that all light-related properties are measured in the
rest-frame SDSS $r$-band~\citep{Stoughton_et_al.(2002)}. All quoted
projection-dependent properties are the mean values of the properties
measured from three principal projections (along the $x$, $y$ and $z$
axes) of the simulation box. A simple semi-analytical dust-attenuation
model which takes dust extinction and scatter into account, as adopted
by~\citet{Xu_et_al.(2017)}, is implemented for the calculations in
this work. We also note that the public online catalogue already contains
many properties which have similar but subtly different
definitions. We specifically use the above-defined properties for this
study, having cross-checked our results also against the properties in
the public catalogue.

\subsection{Galaxy sample selection}
\label{sec:selection}
The dynamically hot stellar component of a galaxy can in principle
build up throughout history and across all galaxy mass scales,
provided gravitational instabilities emerge that cause angular
momentum loss. This is how, in a nutshell, the spheroidal/bulge
components of galaxies and elliptical galaxies in general form. In
this study, we aim at understanding ``what'' the key processes are and
``how'' they {\it start} effectively heating up galaxies while they
are still actively star-forming and retain disk-dominant morphologies.
We focus on systems below a certain galaxy mass scale, such that
further complications due to AGN feedback, which can significantly
affect star-formation cycles, can largely be avoided (e.g. \citealt{Dekel_et_al.(2008)}).  Below, we
present the detailed selection criteria adopted for galaxy samples in
this study.

To start with, we select all central galaxies (excluding all satellite
galaxies) which have stellar masses
$M_{\ast}\,\geqslant\,5\times 10^{9}\,\mathrm{M_{\odot}}$ (below which
galaxies may not have enough particles and not be sufficiently
resolved), where $M_{\ast}$ is the total stellar mass calculated
within a radius of 30 kpc from the galaxy center (in order to exclude
diffuse stellar components,
e.g. \citealt{Schaye_et_al.(2015),Pillepich_et_al.(2018b)}).  We
employ a combination of two criteria to classify star-forming and
disk-dominant galaxies (late-type galaxies), as well as their
spheroidal-shaped counterparts with ceased star-formation (early-type
galaxies):
\begin{enumerate}
\item Similar to the practice of \citet{Genel_et_al.(2018)} and
  \citet{Lu_et_al.(2020)}, we classify galaxies by quantifying their
  distance from the ridge of star-forming main-sequence galaxies. The
  ridge of main-sequence galaxies is defined as the mean sSFR of
  galaxies with $M_{\ast}<10^{10.5}\,\mathrm{M_{\odot}}$, which yields
  $\langle\log\,\mathrm{sSFR/Gyr^{-1}}\rangle\approx -1.0$ at
  $z=0$. Galaxies are classified to be (1) star-forming (late-type
  galaxies) if their sSFR is larger than $0.5\,\rm dex$ below the
  ridge, i.e. $\log\,\mathrm{sSFR/Gyr^{-1}}\geqslant -1.5$ and (2)
  quenched (early-type galaxies) if their sSFR is less than $1\,\rm
  dex$ below the ridge, i.e. $\log\,\mathrm{sSFR/Gyr^{-1}}\leqslant
  -2$.
  
\item A late-type galaxy is also required to have
  $L_{\rm dev}/L_{\rm tot}<0.5$\footnote{{We note that
      in observations, galaxies with pseudo bulges are better
      described by a S{\'e}rsic (rather than a de Vaucouleurs) and an
      exponential profile and the median value of $\mathrm{B/T}$ for
      S0 galaxies is about 0.38. But in this work, we only use this
      rough criterion to select galaxies with thin (disk)
      morphologies.}}, i.e.~a larger luminosity fraction of disk than
  bulge. An early-type galaxy is required to meet the opposite
  criterion.
\end{enumerate} 

\begin{figure*}
\centering
\includegraphics[width=2\columnwidth]{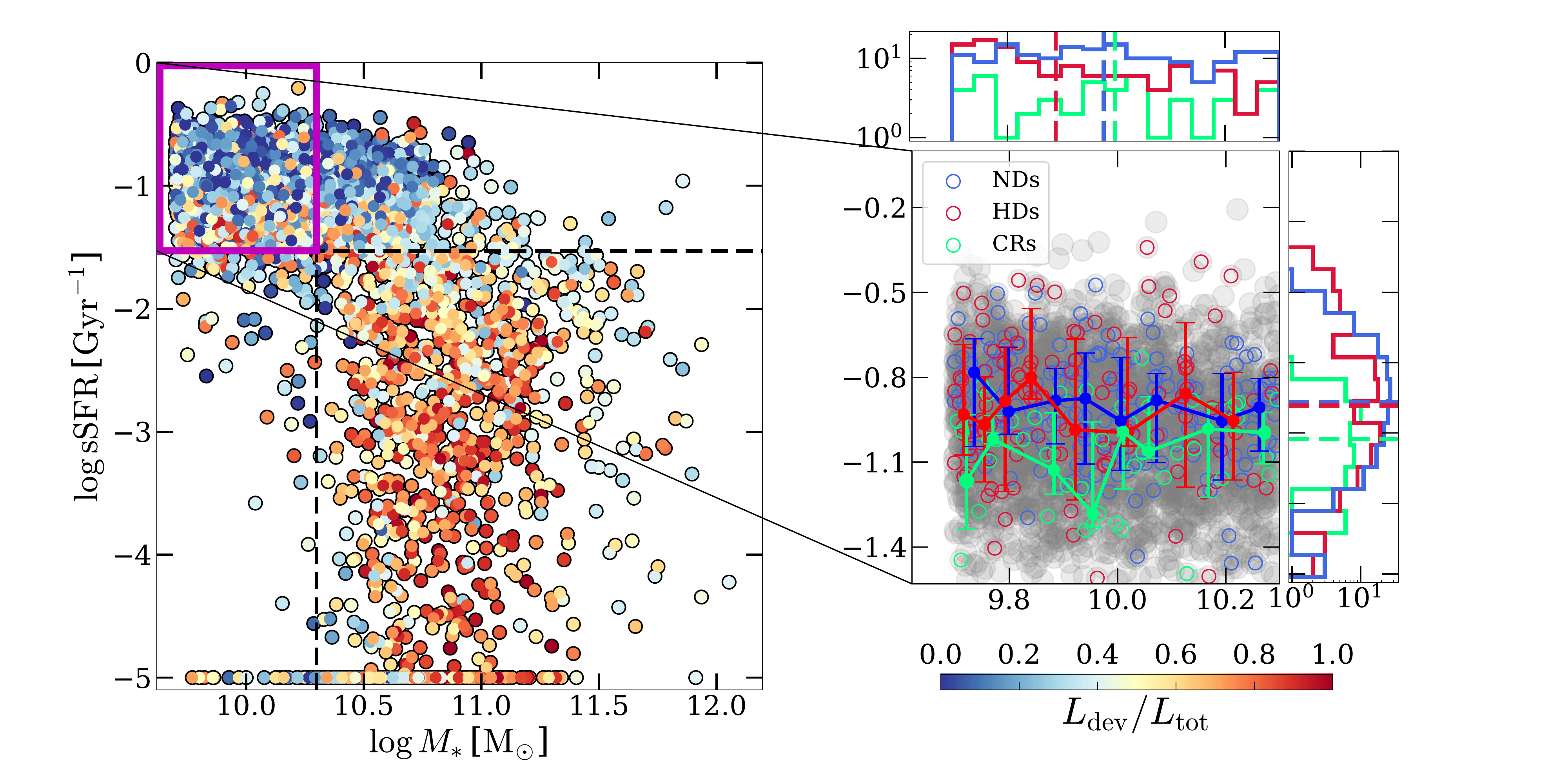}
\caption{Left: specific star-formation rate $\log\,\mathrm{sSFR}$ versus stellar mass $\log\,M_{\ast}$ for all central galaxies in TNG at $z=0$, colour-coded with the bulge component luminosity fraction $L_{\rm dev}/L_{\rm tot}$. The black dashed lines indicate $M_{\ast}=2\times 10^{10}\,\mathrm{M_{\odot}}$ and $\log\,\mathrm{sSFR}/\mathrm{Gyr^{-1}}=-1.5$, respectively. Right: zoom-in distribution of galaxies with $\log\,M_{\ast}/\mathrm{M_{\odot}}\in [9.6,10.3]$ and $\log\,\mathrm{sSFR}/\mathrm{Gyr^{-1}}\in [-1.5,0]$, in which all the galaxies in this region are indicated by grey circles and normal disks (NDs), hot disks (HDs), and counter-rotators (CRs) are indicated with blue, red, and green circles, respectively. The median profile of the three samples are indicated with blue, red, and green lines with error bars indicating the range from the 16th to the 84th percentiles ($1\sigma$).}
\label{fig:ltgs}
\end{figure*}

The two criteria above are applied to all central galaxies at $z=0$
and within the specified stellar mass range. Galaxies that meet both
criteria are classified either into late-types or early-types,
making up $\sim 72\%$ of the total central sample at $z=0$. We confirm
that the $g-r$ colour and S{\'e}rsic index of the two types of galaxies
are naturally well-separated using the criteria above (which can also be
seen in \citealt{Xu_et_al.(2019)}, where similar criteria were
adopted): most late-type galaxies have $g-r$ colour below 0.65 ($\sim$
98\%) and S{\'e}rsic index below 2.5 ($\sim$ 88\%), while most
early-type galaxies have $g-r$ colour over 0.65 ($\sim$ 94\%) and
S{\'e}rsic index over 2.5 ($\sim$ 82\%). We show, in
Fig.~\ref{fig:ltgs} (left panel), the distribution of all central
galaxies in TNG at $z=0$ on the $\log\,\mathrm{sSFR}-\log\,M_{\ast}$
plane, colour-coded with their bulge luminosity fractions
$L_{\rm dev}/L_{\rm tot}$. As can be seen, galaxies with low mass and
high $\mathrm{sSFR}$ tend to have lower bulge fractions, as
theoretically expected.

Among the identified late-type galaxies, we further identify hot disk
(HDs) and counter-rotating (CR) galaxies, both of which, as can be
seen in later sections, are shown to be dynamically hot. HDs are
those where $f_{\rm hot}$ is higher than the mean $f_{\rm hot}$ of the
\textbf{early-type} sample, which is $\sim 0.31$ at $z=0$. CRs are
those whose $f_{\rm ccd}$ is larger than 0.2. For comparison, we also
define (dynamically cold) normal disk galaxies (NDs) whose
$f_{\rm cold}$ is larger than 0.4, above which no HDs and CRs are
observed. In addition, we also adopt a further stellar mass cut at
$M_{\ast}=2\times 10^{10}\,\mathrm{M_{\odot}}$ as the upper limit of
the three selected samples, in order to eliminate the strong
perturbing effects of central AGN activities in these galaxies (e.g. \citealt[fig. 7]{Nelson_et_al.(2018)}).

The former two galaxy types (i.e. HDs and CRs) are in fact minorities in the late-type population, while the latter (i.e. NDs) is the majority. We include all HDs and CRs in our dynamically hot galaxy sample, but only randomly select
NDs to achieve a sample size equal to the sum of the number of HDs and CRs. This results in 118 HDs, 47 CRs and 165 NDs (see Table~\ref{table:table1}). We show the distribution of the three samples in the $\log\,\mathrm{sSFR}-\log\,M_{\ast}$ plane in the right panel of Fig.~\ref{fig:ltgs}, which is a zoom-in figure with galaxies with $\log\,M_{\ast}/\mathrm{M_{\odot}}\in [9.6,10.3]$ and $\log\,\mathrm{sSFR}/\mathrm{Gyr^{-1}}\in [-1.5,0]$. As can be seen, all three sub-populations have roughly similar sSFR, with CRs being slightly lower than the other two. In addition, we have also checked and confirmed that the total halo masses of the three galaxy samples are roughly within the same range.

\begin{figure*}
\centering
\includegraphics[width=2\columnwidth]{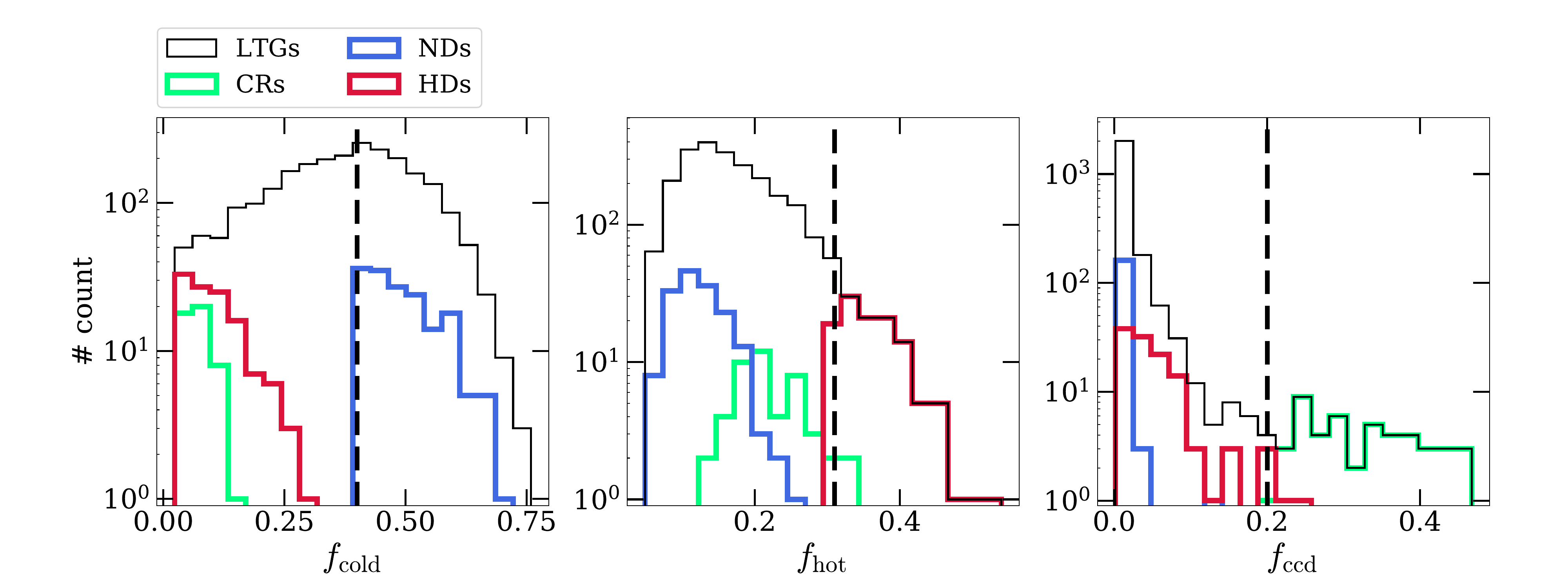}
\caption{Distributions of $f_{\rm cold}$, $f_{\rm hot}$ and $f_{\rm
    ccd}$ (from left to right, see Section~\ref{sec:proptcal} for
  definitions) for late-type galaxies (black), NDs (blue), HDs (red),
  and CRs (green). Black dashed lines represent $f_{\rm cold}=0.4$,
  $f_{\rm hot}=0.31$, and $f_{\rm ccd}=0.2$, which are used to
  classify NDs, HDs, and CRs, respectively.}
\label{fig:forbits}
\end{figure*}

In Fig.~\ref{fig:forbits}, we show the distributions of the three orbital
fractions ($f_{\rm cold}$, $f_{\rm hot}$, and $f_{\rm ccd}$) for
general late-type galaxies and the three selected samples (HDs, CRs
and NDs). As can be seen, the three samples exhibit distinctive
differences in cold, hot and counter-rotating cold orbital fractions:
NDs, which host the highest fractions in $f_{\rm cold}$, also possess
the lowest fractions in $f_{\rm hot}$ and $f_{\rm ccd}$, while HDs
indeed have lower fractions in $f_{\rm cold}$ and $f_{\rm ccd}$, and
CRs have the lowest fractions in $f_{\rm cold}$ and moderate fractions
in $f_{\rm hot}$. This also demonstrates that the three types of galaxies are
intrinsically different from each other.

\begin{table}
  \caption{Number of galaxies in different samples. Col.(1): all $z=0$ central galaxies from
    the IllustrisTNG Simulation; also the parent sample of
    Col.(2)$-$(5). Col.(2): late-type galaxies; Col.(3): selected hot
    disk galaxies (HDs); Col.(4): selected counter-rotating disk
    galaxies (CRs); and Col.(5): control-sample normal disk galaxies
    (NDs). See Section~\ref{sec:selection} for definition and selection
    criteria. } \setlength{\tabcolsep}{5mm}
\begin{tabular}{ccccc}
\hline
\hline
 $N_{\rm central}$ & $N_{\rm late-type}$ & $N^{\rm sel}_{\rm HD}$ & $N^{\rm sel}_{\rm CR}$ & $N^{\rm sel}_{\rm ND}$\\
\hline
5681 & 3177 & 118 & 47 & 165\\
\hline
\end{tabular}
\vspace{2mm}
\label{table:table1}
\end{table}

\section{What do they look like?}
\label{sec:look_like}

\subsection{Morphology of stars and gas}
\label{sec:mor_of_gas_and_star}
\begin{figure*}
\includegraphics[width=2.1\columnwidth]{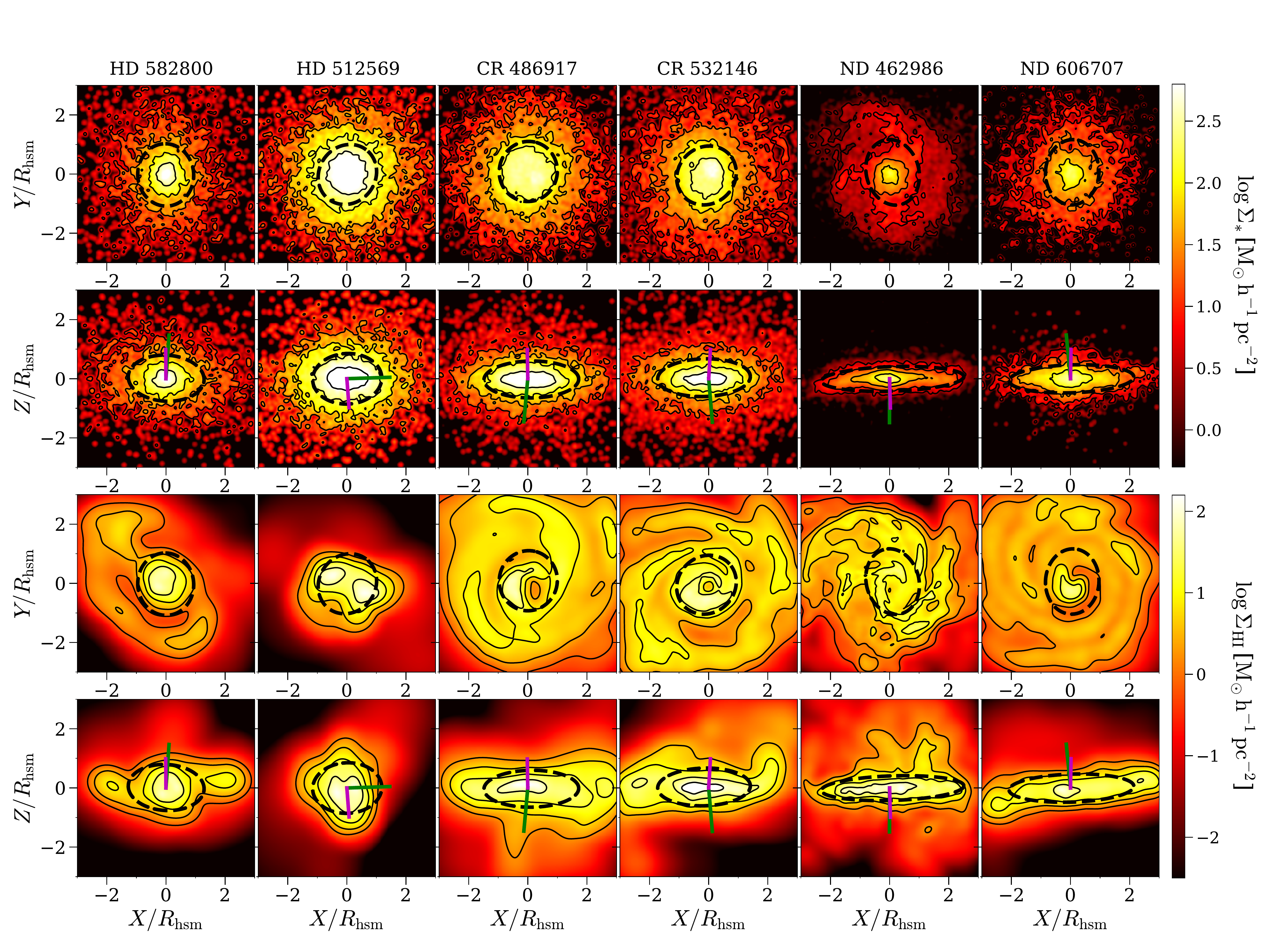}
\caption{Examples of two HDs (ID: 582800, 512569), two CRs (ID: 486917, 532146), and two NDs (ID: 462986, 606707) from left to right. For each example, we show from the top to the bottom rows: \textbf{(1)} the face-on view of the stellar mass density map, \textbf{(2)} the edge-on view of the stellar mass density map, \textbf{(3)} the face-on view of the HI gas surface mass density map, and \textbf{(4)} the edge-on view of the HI gas surface mass density map. The black dashed ellipses indicate the position angle and the ellipticity of the {\it stellar} component for face-on (Row 1) and edge-on (Row 2) views, respectively. Contours of the surface mass densities are shown every 0.5 dex with a set of black curves. In the edge-on maps, the purple and green lines indicate the spin axes of stellar and gaseous components, respectively.}
\label{fig:examples_morphology}
\end{figure*}

To discuss the visually apparent differences in morphologies between
HDs, CRs, and NDs, we first present in
Fig.~\ref{fig:examples_morphology} the stellar and HI mass density
maps from both edge-on and face-on views for typical HDs (the first 2
columns), CRs (the middle 2 columns), and NDs (the latter 2
columns). As can be seen, the two NDs have the flattest stellar and
gaseous disk morphologies with their spin axes well-aligned. In
comparison, however, both CRs have less flat stellar and gaseous
disks, the spins of which exhibit very large misalignment
($\sim 180^{\circ}$); the two HDs possess much thicker stellar disks
and display the least established gaseous disks among all three
types. Interestingly, very often the optical images of NDs and CRs
show clear extended spiral patterns which are absent in the HD samples.

\begin{figure*}
\includegraphics[width=1.8\columnwidth]{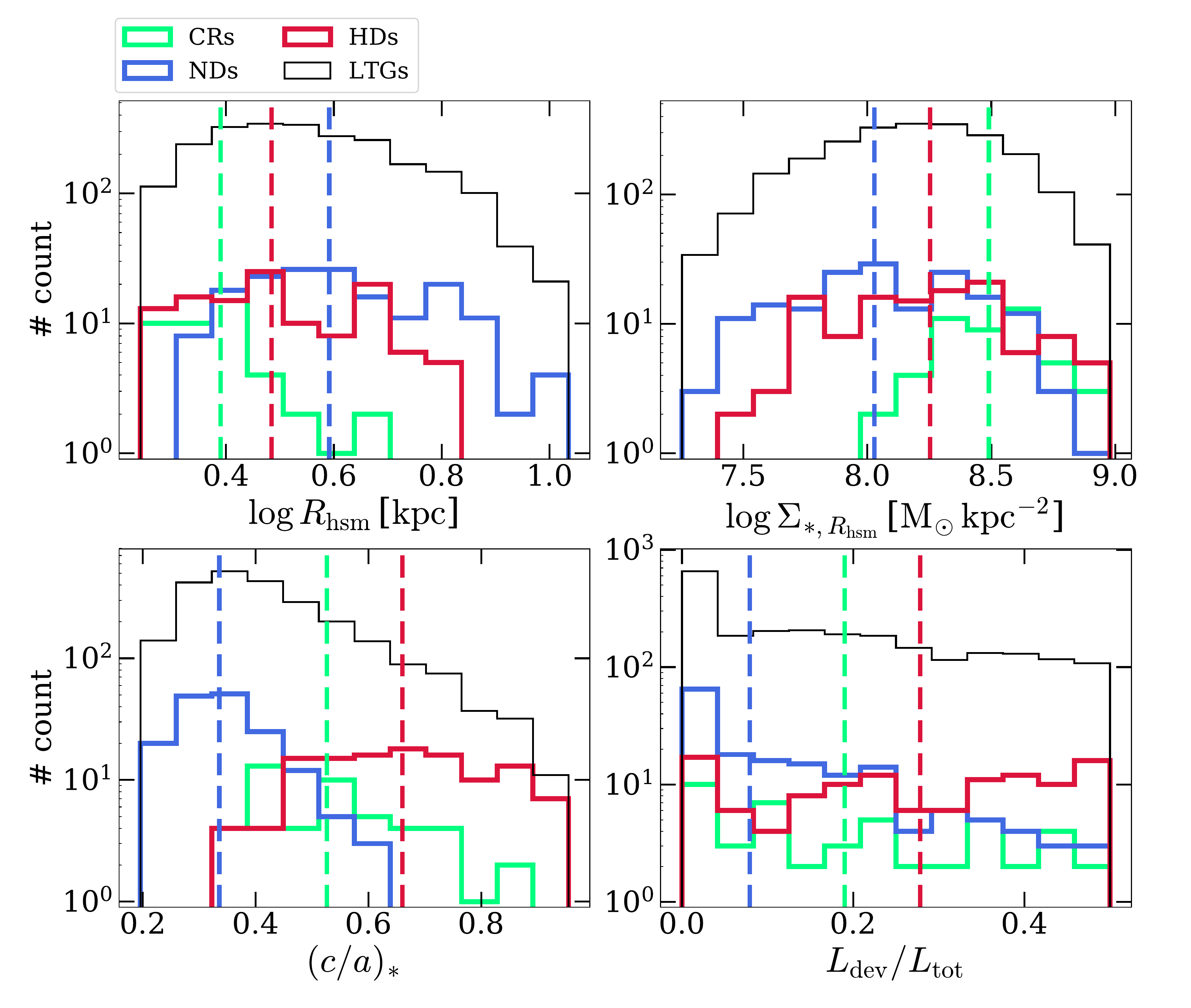}
\caption{Distributions of galaxy size, $\log\,R_{\rm hsm}$ (upper left), surface mass density within $R_{\rm hsm}$, $\log\,\Sigma_{\ast,R_{\rm hsm}}$ (upper right), the shortest-to-longest principal axes ratio of stellar component, $(c/a)_{\ast}$ (bottom left), and de Vaucouleurs to total luminosity ratio, $L_{\rm dev}/L_{\rm tot}$ (bottom right). In each panel, the dashed lines indicate the median values of the corresponding property of NDs (blue), HDs (red), and CRs (green).}
\label{fig:stellar_mor}
\end{figure*}

In Fig.~\ref{fig:stellar_mor} we show the distributions of several
morphology-related key parameters for the stellar components, namely
the half stellar mass radius, $\log\,R_{\rm hsm}$, the surface stellar
mass density within $R_{\rm hsm}$, $\log\,\Sigma_{\ast,R_{\rm hsm}}$,
the shortest-to-longest principal axes ratio of the stellar component,
$(c/a)_{\ast}$, and the luminosity fraction of the de Vaucouleurs
component, $L_{\rm dev}/L_{\rm tot}$. As can be seen, on average NDs
appear to be larger and more disky (smaller $(c/a)_{\ast}$ and $L_{\rm
  dev}/L_{\rm tot}$) with lower surface stellar mass densities, while
HDs and CRs are smaller and relatively speaking more bulge-dominated
(although still having $L_{\rm dev}/L_{\rm tot}<0.5$) with higher
surface stellar mass densities. Specifically, the average sample of
CRs has the smallest sizes (as a result, the highest surface stellar
mass densities) and the average sample of HDs has the largest
$(c/a)_{\ast}$ and $L_{\rm dev}/L_{\rm tot}$ among all three galaxy
types.

\begin{figure*}
\centering
\includegraphics[width=2\columnwidth]{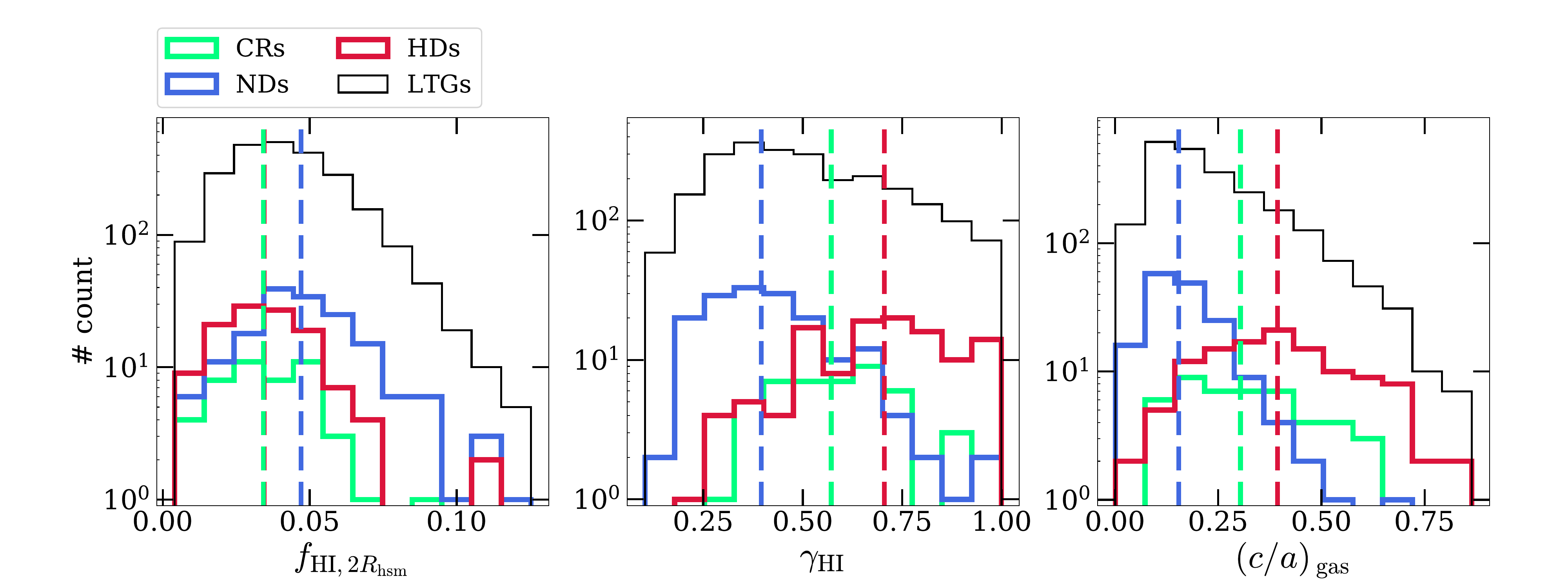}
\caption{Distributions of HI mass fraction within $2R_{\rm hsm}$,
  $f_{\mathrm{HI},2R_{\rm hsm}}$ (left), the ratio of HI mass within
  $R_{\rm hsm}$ and $2R_{\rm hsm}$, $\gamma_{\rm HI}$
  ($\equiv M_{\mathrm{HI},R_{\rm hsm}}/M_{\mathrm{HI},2R_{\rm hsm}}$,
  middle), and the shortest-to-longest principal axes ratio of gas,
  $(c/a)_{\,\rm gas}$ (right). The symbols are the same as in
  Fig.~\ref{fig:stellar_mor}.}
\label{fig:gas_mor}
\end{figure*}

Fig.~\ref{fig:gas_mor} presents the distributions of three
morphological parameters for the gaseous components, namely the
shortest-to-longest principal axes ratio $(c/a)_{\,\rm gas}$, a
concentration parameter of HI gas defined as $\gamma_{\rm HI} \equiv m_{\mathrm{HI},R_{\rm hsm}}/m_{\mathrm{HI},2R_{\rm hsm}}$, and the HI mass fraction $f_{\mathrm{HI},2R_{\rm hsm}}$ calculated within $2R_{\rm hsm}$.  On
average, NDs contain the highest gas fractions $f_{\mathrm{HI},2R_{\rm hsm}}$ confined within the thinnest disks (lowest $(c/a)_{\,\rm gas}$) among all three populations. The gas content of HDs and CRs makes up smaller mass fractions and is distributed within thicker disks in comparison. Interestingly, the latter two cases have higher $\gamma_{\rm HI} $ than the former, suggesting that the HI gas in the dynamically hot disk galaxies
(especially in HDs), although being of low fraction, is more concentrated in the central regions than in their normal disk counterparts, consistent with the examples showcased in Fig.~\ref{fig:examples_morphology}. This is a natural consequence of lower (higher) gas angular momenta in HDs and CRs (NDs), as shall be seen in Section\,\ref{sec:kin_of_gas_and_star}. Besides, the surface HI mass density in IllustrisTNG is roughly consistent with the result shown in \citet{Wang_et_al.(2016)}.

\subsection{Kinematics of stars and gas}

\label{sec:kin_of_gas_and_star}
\begin{figure*}
\centering
\includegraphics[width=2.0\columnwidth]{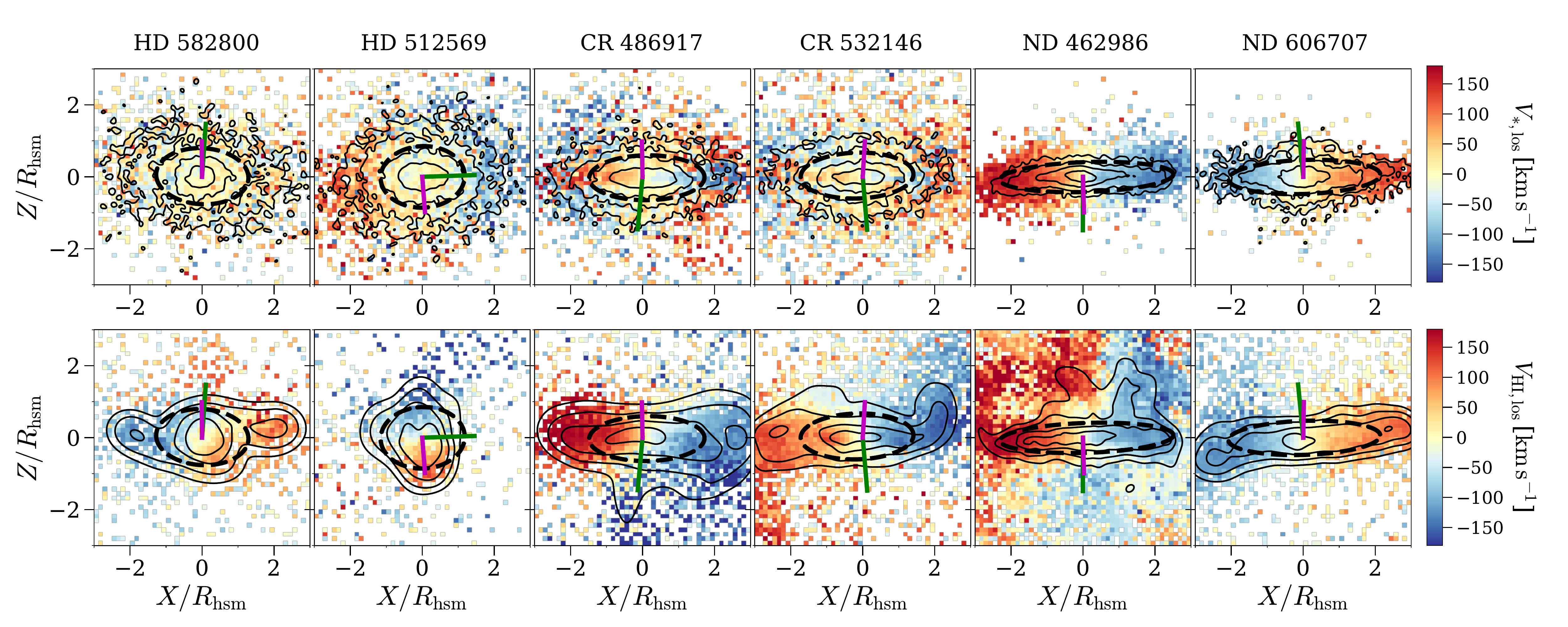}
\caption{Examples of two HDs (ID: 582800, 512569), two CRs (ID: 486917,
  532146), and two NDs (ID: 462986, 606707) from left to right. For each
  example, we show \textbf{(1)} the edge-on view of the line-of-sight
  velocity map of the stellar component (top row) and \textbf{(2)} the edge-on view
  of the line-of-sight velocity map of HI gas (bottom row). In each panel, the
  position angle and the ellipticity of the {\it stellar} component
  from edge-on view are indicated by the black ellipse, with purple
  and green lines indicating the spin axes of the stellar and gaseous
  components, respectively; contours of surface mass density are shown
  every 0.5 dex by a set of black curves.}
\label{fig:examples_kinematics}
\end{figure*}

\begin{figure*}
\centering
\includegraphics[width=2.1\columnwidth]{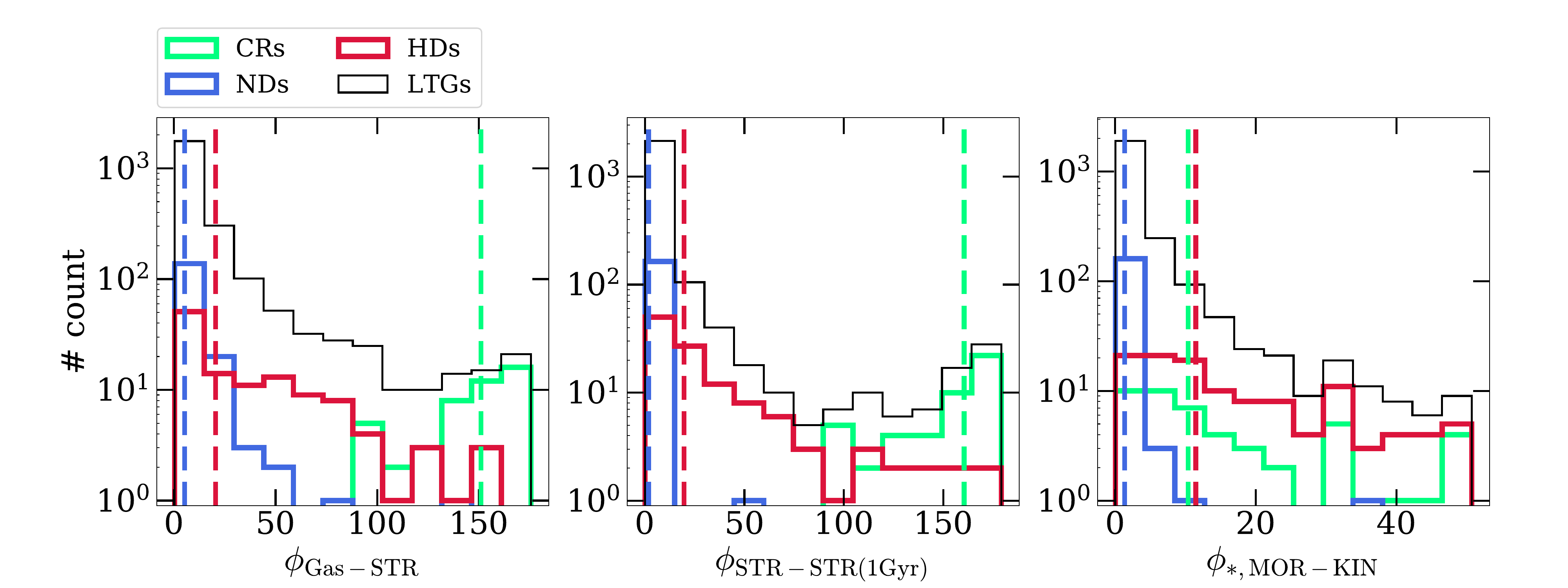}
\caption{Distributions of misalignment angles between \textbf{(1)} the kinematic major axes of gas and stars, $\phi_{\rm Gas-STR}$, \textbf{(2)} the kinematic major axes of stars, and young stars formed in the recent $1\,\mathrm{Gyr}$, $\phi_{\rm STR-STR(1Gyr)}$, and \textbf{(3)} the photometric and kinematic major axes of stars, $\phi_{\ast,\rm MOR-KIN}$ from left to right. The symbols are the same as in Fig.~\ref{fig:stellar_mor}.}
\label{fig:spin_misalignment}
\end{figure*}

Fig.~\ref{fig:examples_kinematics} presents the line-of-sight velocity
maps from the edge-on view of both stars and HI gas for our example galaxies. As can be seen, the stellar and HI disks of the two NDs show a co-planar and
co-rotating pattern, which is absent in HDs and CRs. The two HDs show much weaker rotation in both stellar and gaseous disks. As shall be seen in Fig.~\ref{fig:spin_misalignment}, very often HDs exhibit kinematically and morphologically misaligned stellar and gaseous disks. The CR galaxy samples are selected according to hosting a good fraction of counter-rotating stellar orbits, which is indeed revealed by the stellar kinematic maps of the two example CR galaxies. Interestingly, the HI gas therein is in fact also rotating in the opposite direction relative to the bulk stellar population (as indicated by their markedly different spin directions labelled in the figure). In the following sections, we shall see that this is a natural outcome of their evolution.

\begin{figure*} 
\includegraphics[width=1.8\columnwidth]{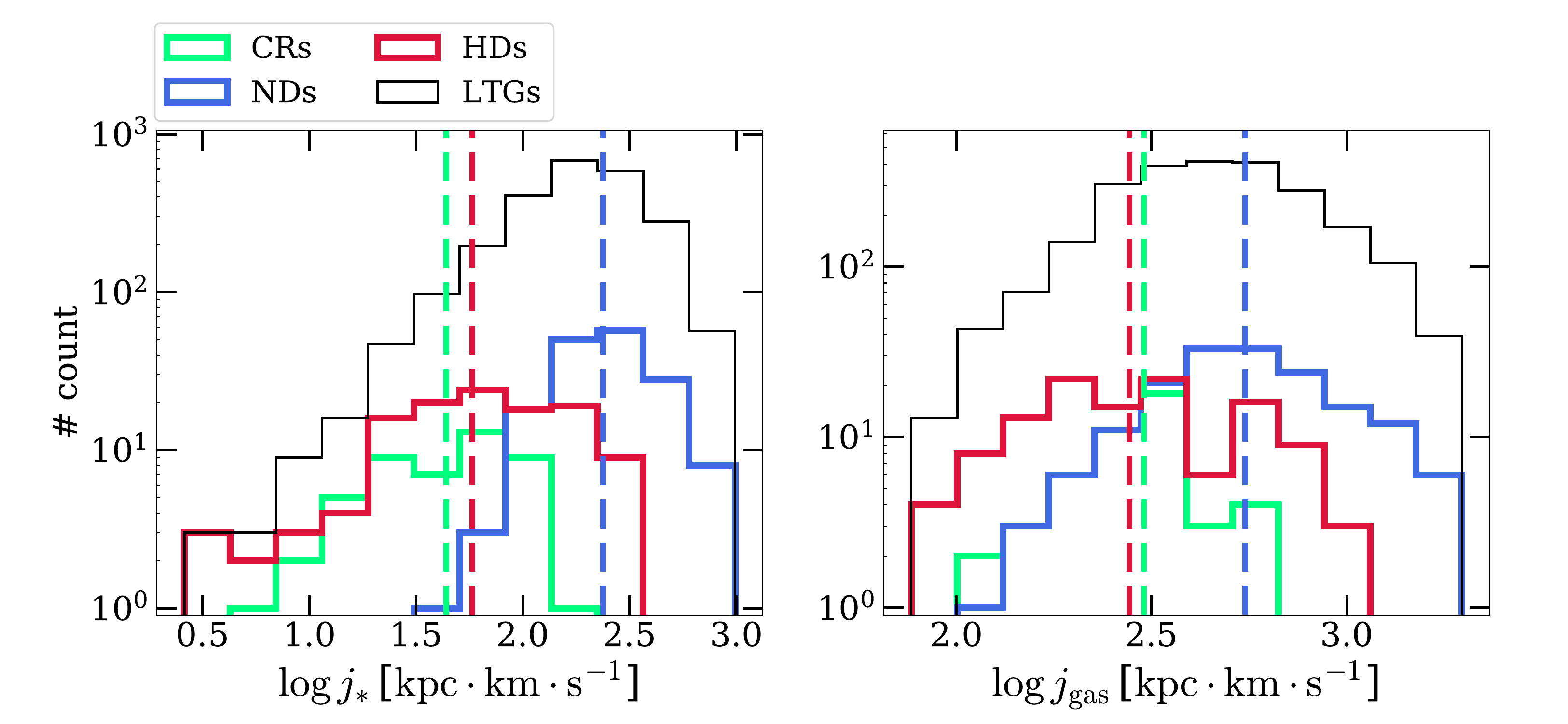}
\caption{Distributions of spin of the stellar component
  $\log\,j_{\ast}$ (left) and the gas component $\log\,j_{\rm gas}$
  (right). The symbols are the same as in Fig.~\ref{fig:stellar_mor}.}
\label{fig:spins}
\end{figure*}

The misalignment between the (bulk) stellar and the gas components is
shown even more clearly in Fig.~\ref{fig:spin_misalignment}, where we
present the distributions of the misalignment angles between
\textbf{(1)} the spin axes of stellar and gaseous components within
$2R_{\rm hsm}$, $\phi_{\rm Gas-STR}$, \textbf{(2)} the stellar
component and stars formed in the recent 1 Gyr within $2R_{\rm hsm}$,
$\phi_{\rm STR-STR(1Gyr)}$, and \textbf{(3)} the shortest principal
axis and the spin axis of the stellar component, $\phi_{\ast,\rm MOR-KIN}$
(from left to right). As shown in the figure, while NDs always have
well aligned stellar and gaseous disks, HDs and CRs exhibit remarkably
large misalignment: the angles $\phi_{\rm Gas-STR}$ and
$\phi_{\rm STR-STR(1Gyr)}$ for most of the HDs occupy the range
between $\sim 0^{\circ}-90^{\circ}$; while the CRs show these two
misalignment angles typically between $\sim 90^{\circ}-180^{\circ}$,
revealing their gaseous disks rather than tracing the youngest stellar
populations, both of which are counter-rotating with respect to the
spin of the bulk (older) stellar populations. Besides, HDs and CRs
expand their $\phi_{\ast,\rm MOR-KIN}$ up to $\sim 50^{\circ}$, while
NDs rarely have $\phi_{\ast,\rm MOR-KIN}$ over $10^{\circ}$. In
Fig.~\ref{fig:spins} we show the distributions of the stellar and
gaseous spins for the three galaxy samples. In general, NDs possess
the highest spins while HDs and CRs have weaker co-rotation, as is
also demonstrated in Fig.~\ref{fig:examples_kinematics} for example
galaxies.

\begin{figure*}
\includegraphics[width=1.8\columnwidth]{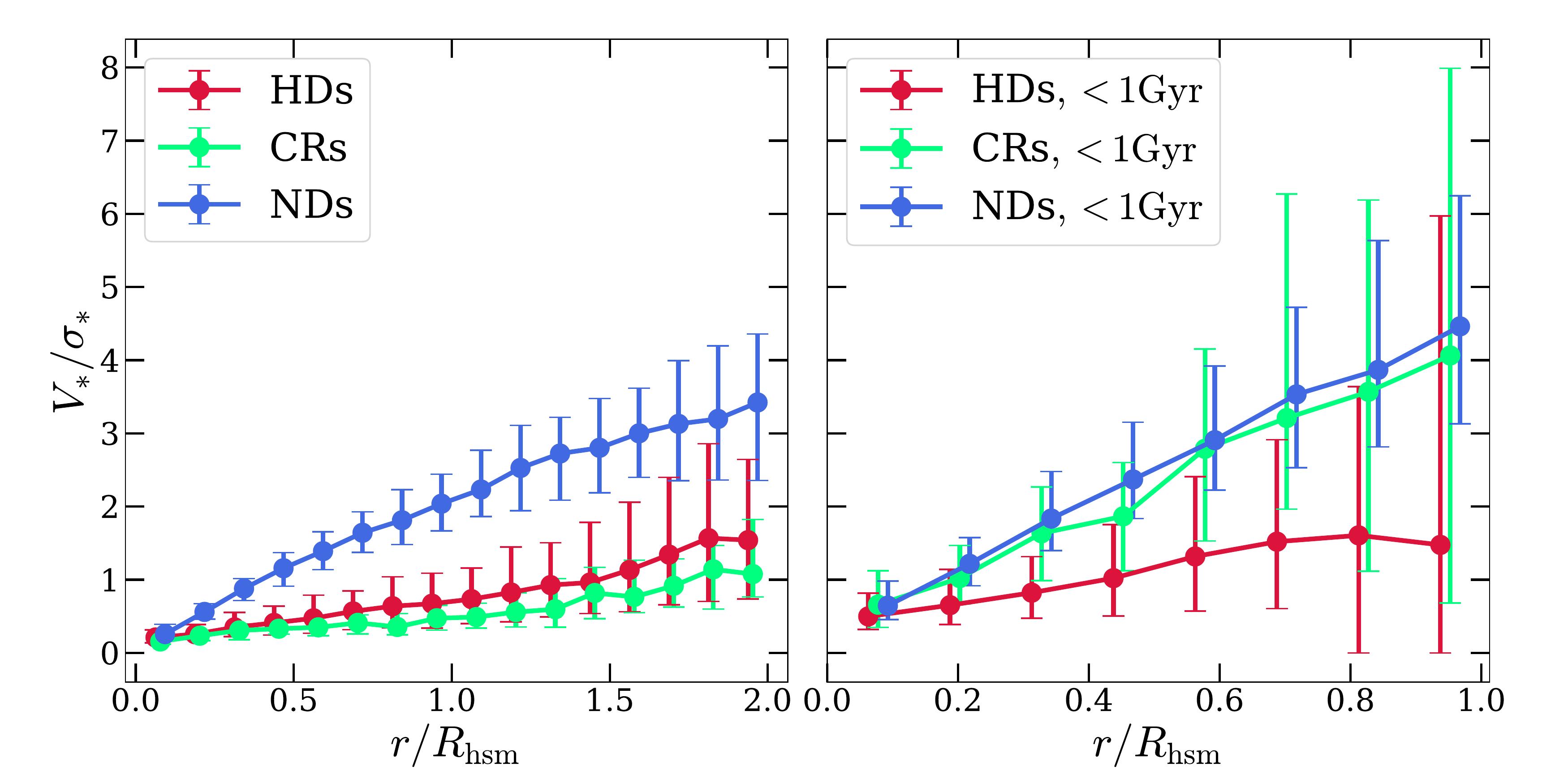}
\caption{Radial $V_{\ast}/\sigma_{\ast}$ profiles for HDs (red), CRs
  (green), and NDs (blue), where $V_{\ast}$ is the line-of-sight
  velocity and $\sigma_{\ast}$ is the dispersion. For each galaxy,
  both $V_{\ast}$ and $\sigma_{\ast}$ are calculated in a slit with
  its width being 0.5$R_{\rm hsm}$ along the major axis from the
  edge-on view of the galaxy. The left panel shows the
  $V_{\ast}/\sigma_{\ast}$ profiles of all stars in the slit and the
  right panel shows the $V_{\ast}/\sigma_{\ast}$ profiles of stars
  formed in the recent 1$\,\mathrm{Gyr}$ (shared $y$-axis with the left
  panel). The error bars indicate the range from the 16th to the 84th
  percentiles (1$\sigma$).}
\label{fig:v2sigma}
\end{figure*}

Observationally, the stellar spin of a galaxy is difficult to measure
directly. A good observational indicator is $V_{\ast}/\sigma_{\ast}$,
where $V_{\ast}$ is the line-of-sight velocity and $\sigma_{\ast}$ is
the dispersion, both of which can be obtained directly from
spectroscopic observations. In Fig.~\ref{fig:v2sigma} we show the
average radial profile of $V_{\ast}/\sigma_{\ast}$ for the three
galaxy samples. Here the $V_{\ast}/\sigma_{\ast}$ profile of each
galaxy is calculated in a long slit along the semi-major axis from the
galaxy's edge-on view, with a width of 0.5$R_{\rm hsm}$. As can be
seen in the left panel, while NDs show steeply increasing
$V_{\ast}/\sigma_{\ast}$ profiles towards larger radii with
$V_{\ast}/\sigma_{\ast} \sim 3.5$ at $r=2R_{\rm hsm}$, HDs and CRs
both show flatter $V_{\ast}/\sigma_{\ast}$ profiles with
$V_{\ast}/\sigma_{\ast} \sim 2$ and $\sim 1.3$ at $r=2R_{\rm hsm}$,
respectively, again indicating their weaker co-rotation in comparison.
On the right, the $V_{\ast}/\sigma_{\ast}$ profiles of young stars
(formed in the recent $1\,\mathrm{Gyr}$) are presented. For HDs, the
$V_{\ast}/\sigma_{\ast}$ profile remains flatter with a ratio
$\sim 1.2$ at $r=R_{\rm hsm}$. Note that strong rotation features as
indicated by large $V_{\ast}/\sigma_{\ast}$ are clearly seen both for
NDs and CRs, with a ratio of $\sim 4$ at $r=R_{\rm hsm}$.

\begin{figure*}
\centering
\includegraphics[width=1.8\columnwidth]{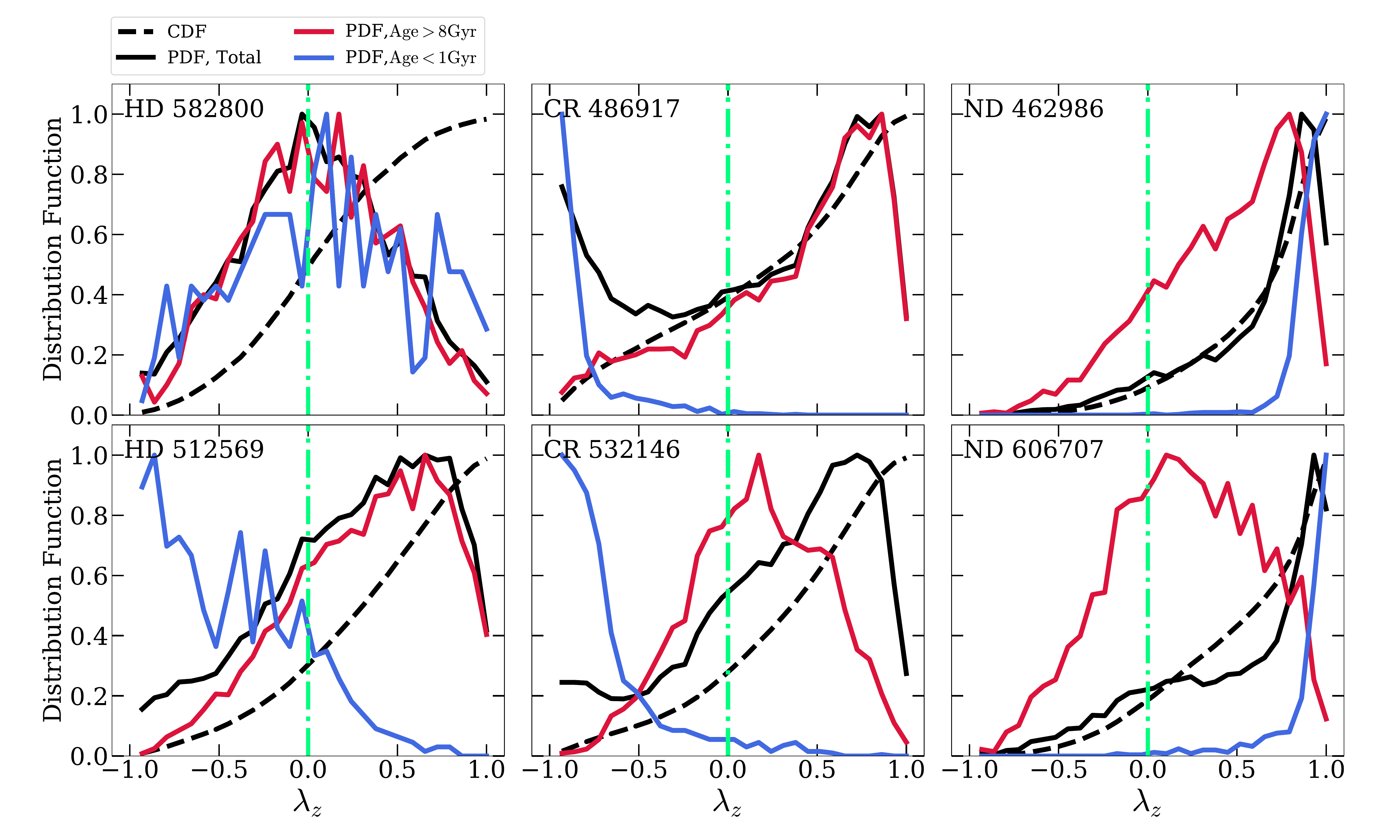}
\caption{The distribution of circularity $\lambda_z$ ($\equiv L_{\rm z}/J_{\rm c}$)
  of two example HDs (ID: 582800, 512569), two example CRs (ID:
  486917, 532146), and two example NDs (ID: 462986, 606707) from left
  to right. In each panel, the black dashed curve represents the CDF
  of $L_{\rm z}/J_{\rm c}$ and the black solid curve shows the PDF of all stars
  within $2R_{\rm hsm}$; the PDFs of stars formed 8 Gyr ago and within
  the recent 1 Gyr (both are within $2R_{\rm hsm}$) are shown by red
  and blue curves, respectively. The green dotted-dashed line
  indicates $\lambda_z=0$.}
\label{fig:examples_circularity}
\end{figure*}

The above described behaviour of $V_{\ast}/\sigma_{\ast}$ among the
three galaxy samples can be better understood and demonstrated from
the perspective of stellar orbits. In
Fig.~\ref{fig:examples_circularity}, we present the distribution of
circularity $\lambda_z$ ($\equiv L_{\rm z}/J_{\rm c}$, see definition in
Section~\ref{sec:proptcal}) for our example galaxies. As shown in the
figure, the total (black solid) circularity distributions of NDs show clear
peaks towards $\lambda_z\sim 1$, while CRs also exhibit an increasing
tail towards $\lambda_z\sim -1$, indicating the existence of a good
fraction of counter-rotating stellar population. Unlike the two types
above, HDs host a significant fraction of low-angular momentum
($\lambda_z\sim 0$) orbits. When considering only stars that formed
over 8 Gyrs ago, their circularity distributions (red) can be
dominated by either a low-angular momentum component (with $\lambda_z$
peaking close to zero) or high-angular momentum circular orbits
already existing by then. Stars that formed less than 1 Gyrs ago may
exhibit different orbital distributions (blue) from those who have
formed earlier-on. Newly-formed stars in NDs have circularities
significantly peaking at $\lambda_z\sim 1$, indicating a strongly
established co-rotating disk at the current epoch, while those in CRs
clearly occupy counter-rotating orbits ($\lambda_z \sim
-1$). Newly-formed stars in some of the HDs behave quite similarly to
those in CRs (e.g., HD 512569). Such HDs in fact used to be CRs at
earlier times (see next section for more details). There are also HDs
which have always been dominated by hot orbits ($\lambda_z\sim 0$,
e.g., HD 582800) throughout their histories.

It is worth noting that when we combine all above-mentioned kinematic
properties of gas and stars, we find that in CR galaxies, newly formed
stars in fact follow circular motions, as a consequence of being
formed in gaseous disks that are counter rotating with respect to the
bulk (older) stellar populations in these systems. This was also
reported in~\citet{Starkenburg_et_al.(2019)}, who studied possible
formation scenarios of low-mass counter-rotating galaxies in the
Illustris
simulation~\citep{Vogelsberger_et_al.(2013),Vogelsberger_et_al.(2014a),Vogelsberger_et_al.(2014b),Genel_et_al.(2014),Nelson_et_al.(2015)}. In
Section~\ref{sec:how_form}, we present some alternative origins added
to their formation picture.

\subsection{Star-formation activities}
\label{sec:SF_activity}
\begin{figure*}
\centering
\includegraphics[width=2.1\columnwidth]{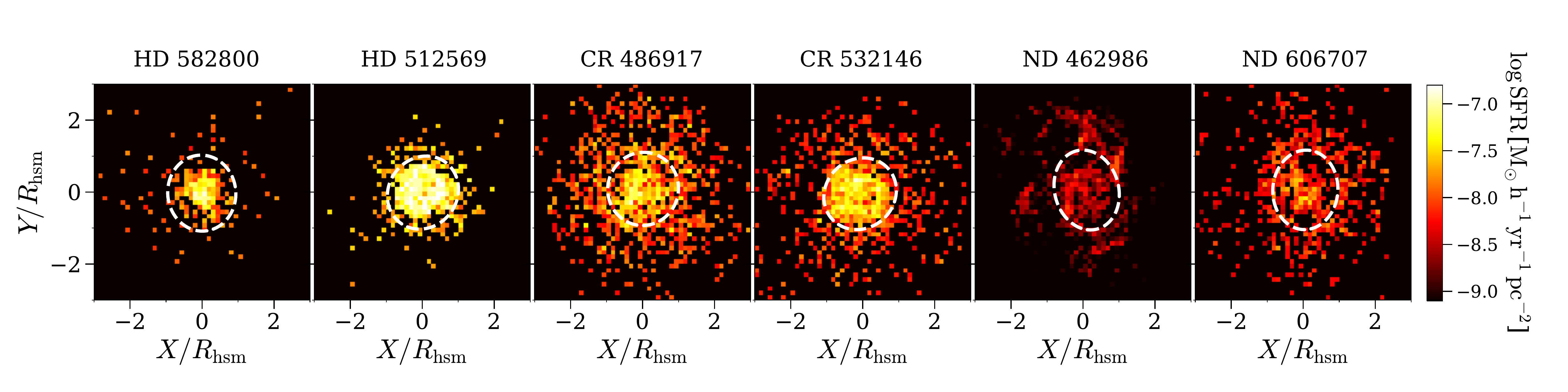}
\caption{Examples of two HDs (ID: 582800, 512569), two CRs (ID:
  486917, 532146), and two NDs (ID: 462986, 606707) from left to
  right. For each example, we show the face-on map of its
  star-formation rate density in the recent $1\,\mathrm{Gyr}$, with
  the white ellipse indicating the position angle and ellipticity
  obtained from the face-on view of the example.}
\label{fig:examples_SFR}
\end{figure*}

In this section, we show the comparison of star-formation activities
among the three galaxy samples. We first present the star-formation-rate
maps for the six example galaxies in Fig.~\ref{fig:examples_SFR}. 
As can be seen, the two NDs typically have extended star-formation 
activities (out to $\sim 3R_{\rm hsm}$), while star-formation  
in the two HDs is strongly constrained within the central regions; 
the two CRs exhibit both a core and fairly extended star-forming 
distributions. We attribute this central concentration in star-formation 
in these dynamically hot systems to the lower gas angular momenta 
therein (see Fig.~\ref{fig:spins}).

\begin{figure*}
\includegraphics[width=1.8\columnwidth]{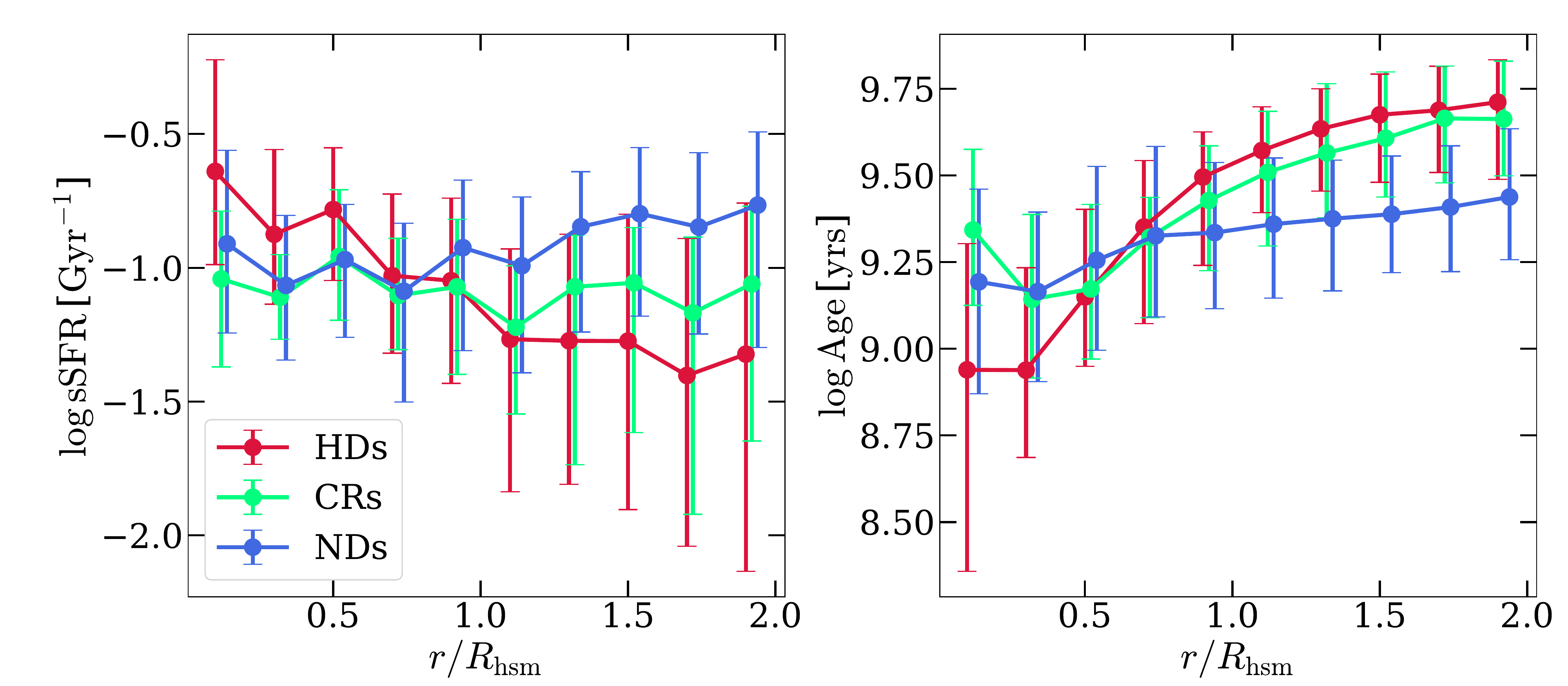}
\caption{Radial profiles of the specific star-formation rate
  ($\log\mathrm{sSFR}$, left) and age ($\log\mathrm{Age}$,
  right). HDs, CRs, and NDs are represented by red, green, and blue
  lines, respectively. The error bars indicate the range from the 16th
  to the 84th percentiles ($1\sigma$).}
\label{fig:sp_profiles}
\end{figure*}

To better illustrate the differences in the spatial distributions of
star-formation activities among the three galaxy samples, we show in
Fig.~\ref{fig:sp_profiles} the specific star-formation-rate (sSFR) and stellar age profiles as
a function of radius out to $2R_{\rm hsm}$ from the galaxy center.  As
can be seen, HDs (red) have on average the most concentrated star
formation, which is also the strongest among the three galaxy samples,
as far as the central regions are concerned. As a consequence, the
central stellar populations therein are significantly younger than the
outskirts as well as those in the ND and CR counterparts. In contrast,
NDs (blue) have fairly flat/extended\footnote{We note that the ND
  samples in this study are selected to be of lower-masses and have
  not established their bulge components therefore the age
  distributions in these systems do not yet show a significantly older
  population in the central regions as expected from the
  ``inside-out'' growth model.} distributions in both sSFR and stellar
age. In comparison, NDs have the highest sSFR and the youngest stellar
ages at larger disk radii among all three galaxy samples. On the other
hand, CR galaxies (green) bear some resemblances to both HDs and NDs:
centrally-concentrated star-formation is clearly seen, while at larger
radii, sSFR therein does not as quickly drop as in HDs, but is not yet
as high as in NDs. We discuss the reasons for these behaviours in
Section~\ref{sec:how_form}.

\section{How do they form? }

\label{sec:how_form}
\citet{Starkenburg_et_al.(2019)} investigated the origin of gas-star
counter-rotating (star-forming) systems using the Illustris
simulation, and found that the formation is firstly associated with a
major episode of gas removal through two effective channels, namely
AGN feedback and fly-by passages (identified as being satellite
galaxies prior to counter-rotating), followed by new gas accretion in
anti-aligned directions. In this study, in order to understand the
origin of all three selected galaxy samples in the IllustrisTNG
simulation, we make use of the {\sc sublink}
tree~\citep{Rodriguez-Gomez_et_al.(2015)} provided by the IllustrisTNG
team to trace the entire evolution of the simulated galaxies up to
$z=2$. This allows us to examine the connections between their
evolutionary origin, their star-forming and AGN feedback activities,
their gas content variations, their stellar and gas spin evolution, as
well as their galaxy merger histories. We find that this last aspect
plays a crucial role in shaping the morphological and kinematic
evolution of the galaxies examined in this study.

Apart from the properties of the investigated galaxies along their
main evolutionary branches, we also employ three quantities to
characterize the mergers the galaxies experienced: merger orbit angle
$\theta_{\rm orb}$, merger mass ratio $\mu$, and merger gas fraction
$f_{\rm g}$. Here $\theta_{\rm orb}$ is defined as the angle between
the orbital angular momentum of the in-coming galaxy (measured at the
time when the galaxy has the highest stellar mass across its evolution
history) and the spin of the host galaxy at that snapshot. Thus,
$\theta_{\rm orb}>90^{\circ}$ means that the orbital spin of the
in-coming galaxy is counter-rotating with respect to the spin of the
host galaxy. The merger mass ratio $\mu$ is defined as
\begin{equation}
\label{eq:mu}
   \mu = \frac{M_{\ast,\rm incoming}}{M_{\ast,\rm host}}, 
\end{equation}
where $M_{\ast,\rm incoming}$ and $M_{\ast,\rm host}$ are the stellar
masses of the in-coming galaxy and the host galaxy within
$2R_{\rm hsm}$, respectively. The merger gas fraction is defined as
\begin{equation}
\label{eq:fg}
    f_{\rm g} = \frac{M_{\rm gas,incoming}+M_{\rm gas,host}}{M_{\rm total,incoming}+M_{\rm total,host}},
\end{equation}
where $M_{\rm gas,incoming}$ and $M_{\rm gas,host}$ are gas masses of
the in-coming galaxy and the host galaxy, while $M_{\rm total,incoming}$
and $M_{\rm total,host}$ are the total masses (including stellar masses and dark matter masses) of the two galaxies,
respectively. All these masses are measured within a 3D aperture of
radius equal to $2R_{\rm hsm}$.

\subsection{Two example ND galaxies}

\begin{figure*}
\centering
\includegraphics[width=1\columnwidth]{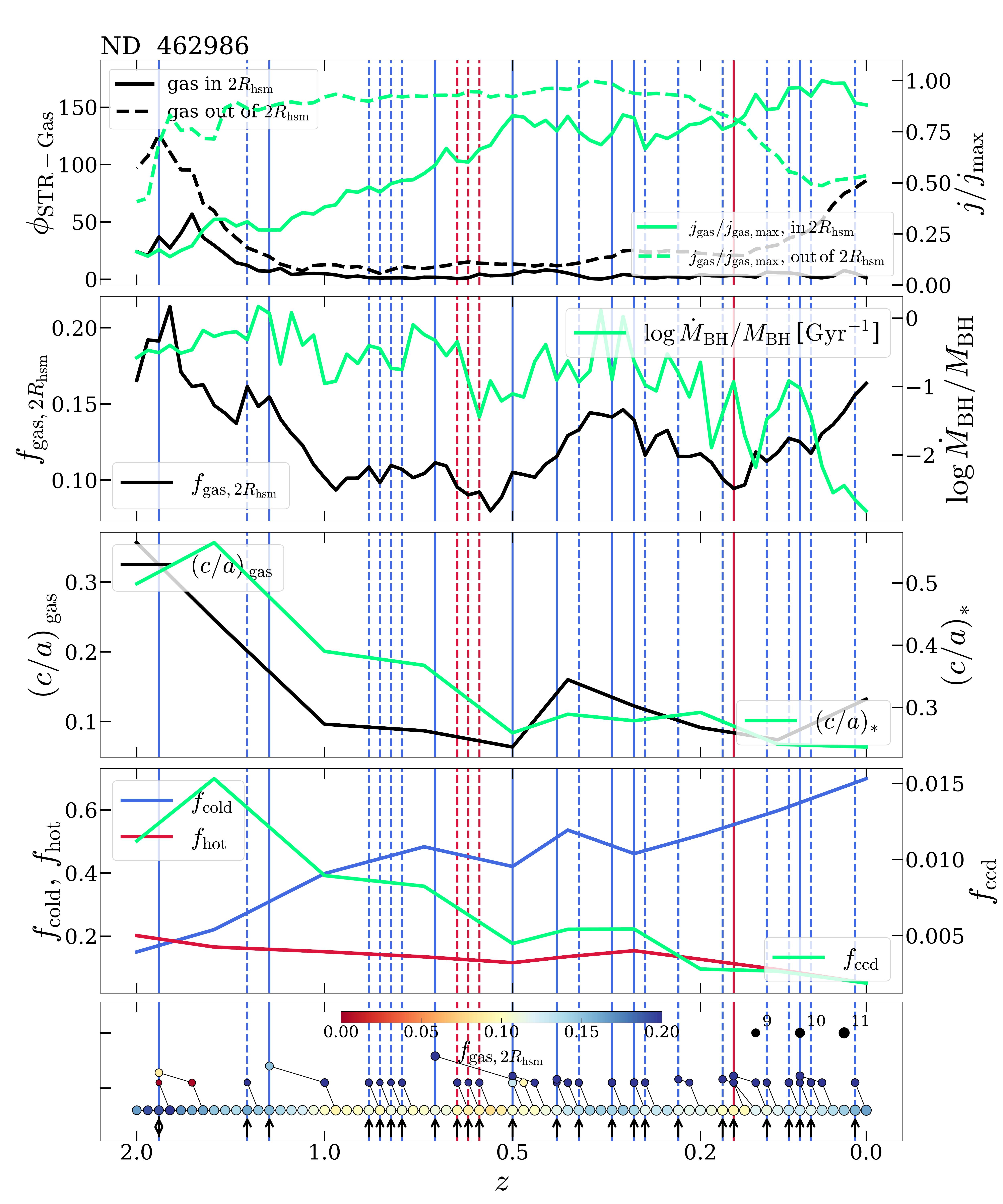}
\includegraphics[width=1\columnwidth]{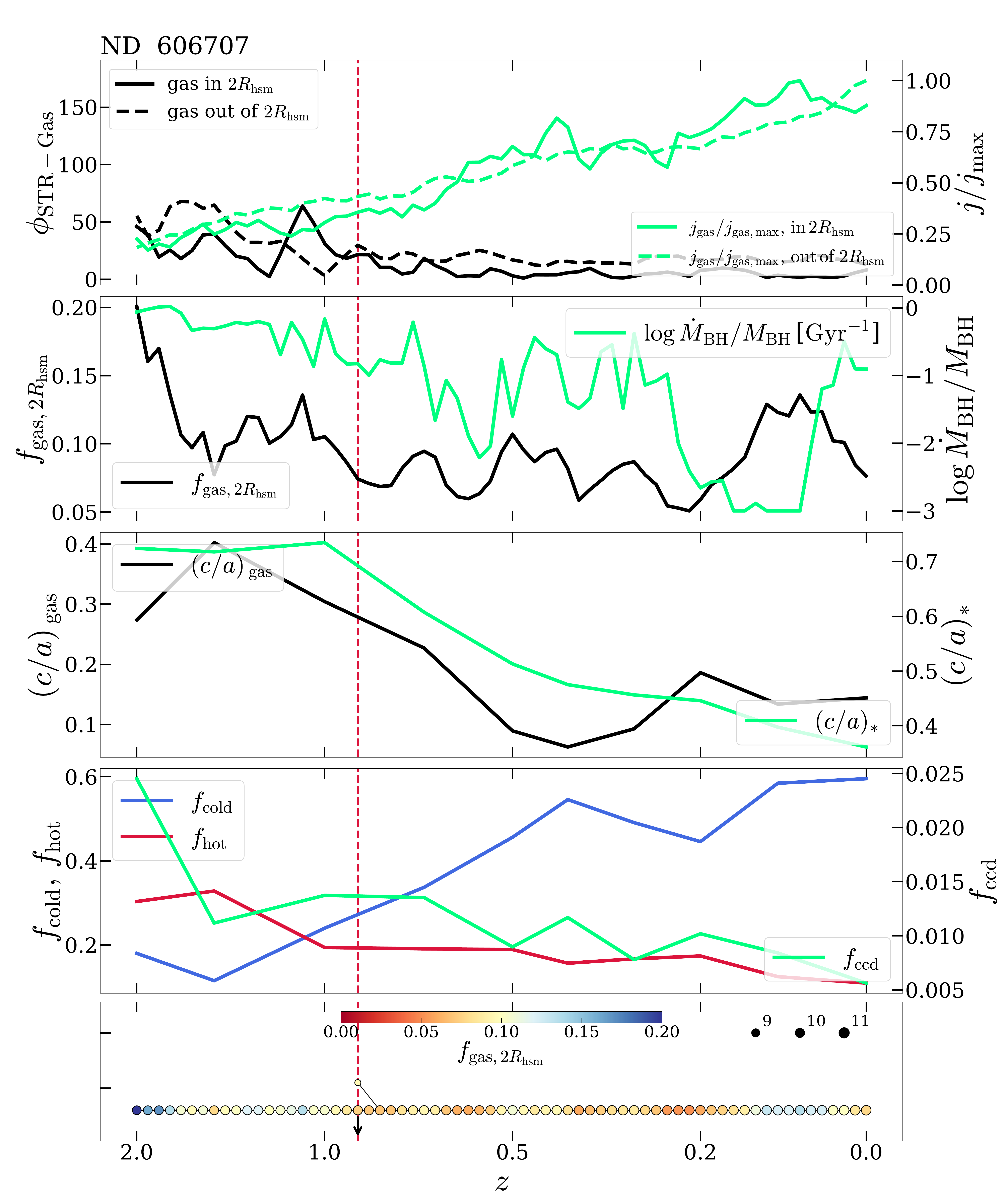}
\caption{The redshift evolution ($0\,\leqslant\,z\,\leqslant\,2$) of
  properties for the two ND examples (IDs: 462986, 606707). The
  properties are \textbf{(1)} the spin axes misalignment of gas within
  (black solid) and out of 2$R_{\rm hsm}$ (black dashed) with respect
  to stars within $2R_{\rm hsm}$, $\phi_{\rm STR-Gas}$ (1st panel,
  left Y-axis); \textbf{(2)} the spin of gas within (green solid) and out of
  2$R_{\rm hsm}$ (green dashed), $j_{\rm gas}/j_{\rm gas,max}$ (normalized to its
  maximum in the history, 1st panel, right Y-axis); \textbf{(3)} gas fraction
  within 2$R_{\rm hsm}$, $f_{\mathrm{gas},2R_{\rm hsm}}$ (black solid,
  2nd panel, left Y-axis); \textbf{(4)} black hole accretion rate,
  $\log\,\dot{M}_{\rm BH}/M_{\rm BH}$ (green solid, 2nd panel, right Y-axis);
  \textbf{(5)} the shortest-to-longest axis ratio for gas,
  $(c/a)_{\,\rm gas}$ (black solid, 3rd panel, left Y-axis); \textbf{(6)} the
  shortest-to-longest axis ratio for the stellar component, $(c/a)_{\ast}$
  (green solid, 3rd panel, right Y-axis); and \textbf{(7)} 3 types of orbital
  fractions (4th panel): $f_{\rm cold}$ (blue solid, left Y-axis),
  $f_{\rm hot}$ (red solid, left Y-axis), and $f_{\rm ccd}$ (green solid,
  right Y-axis). For each example, we show its merger history in the bottom
  panel, with circle sizes indicating the stellar mass within
  $2R_{\rm hsm}$ and colours indicating the gas fraction within
  2$R_{\rm hsm}$ of galaxies at different redshifts. For each merger
  event, we show both the in-coming galaxy and its progenitor at the
  snapshot when it reaches its maximum stellar mass. In each panel,
  each merger event is indicated by a vertical line with its colour
  encoding the merger being wet ($f_{\rm g}\,\geqslant\,0.1$, blue)
  or dry ($f_{\rm g}<0.1$, red) and line style marking the merger
  mass ratio ($\mu<0.01$: dashed; $\mu\,\geqslant\,0.01$: solid). In
  the bottom of the merger tree, the arrows represent the orbital
  angle of the in-coming galaxy with respect to the spin of the host
  galaxy, $\theta_{\rm orb}$ ($\theta_{\rm orb}<90^{\circ}$: up arrow;
  $\theta_{\rm orb}\,\geqslant\,90^{\circ}$: down arrow). See
  Section~\ref{sec:how_form} for definitions of $f_{\rm g}$, $\mu$,
  and $\theta_{\rm orb}$.}
\label{fig:evol_ND}
\end{figure*}

Fig.~\ref{fig:evol_ND} shows the evolution of two example NDs taken
from our sample. As can be seen, the angular momenta of both stellar
and gas components grow steadily. The misalignment between the gas and
stellar spins decreases, and both the stellar and gas disks become
thinner and thinner as the dynamically cold stellar orbits build
up. Galaxy ND 462986 has experienced very frequent gas-rich mergers
while Galaxy ND 606707 has undergone only one recorded merger, but
both have successfully established their disk morphology and
rotational kinematics with active star-formation at the current epoch.

\subsection{Two example HD galaxies}

\begin{figure*}
\centering
\includegraphics[width=1\columnwidth]{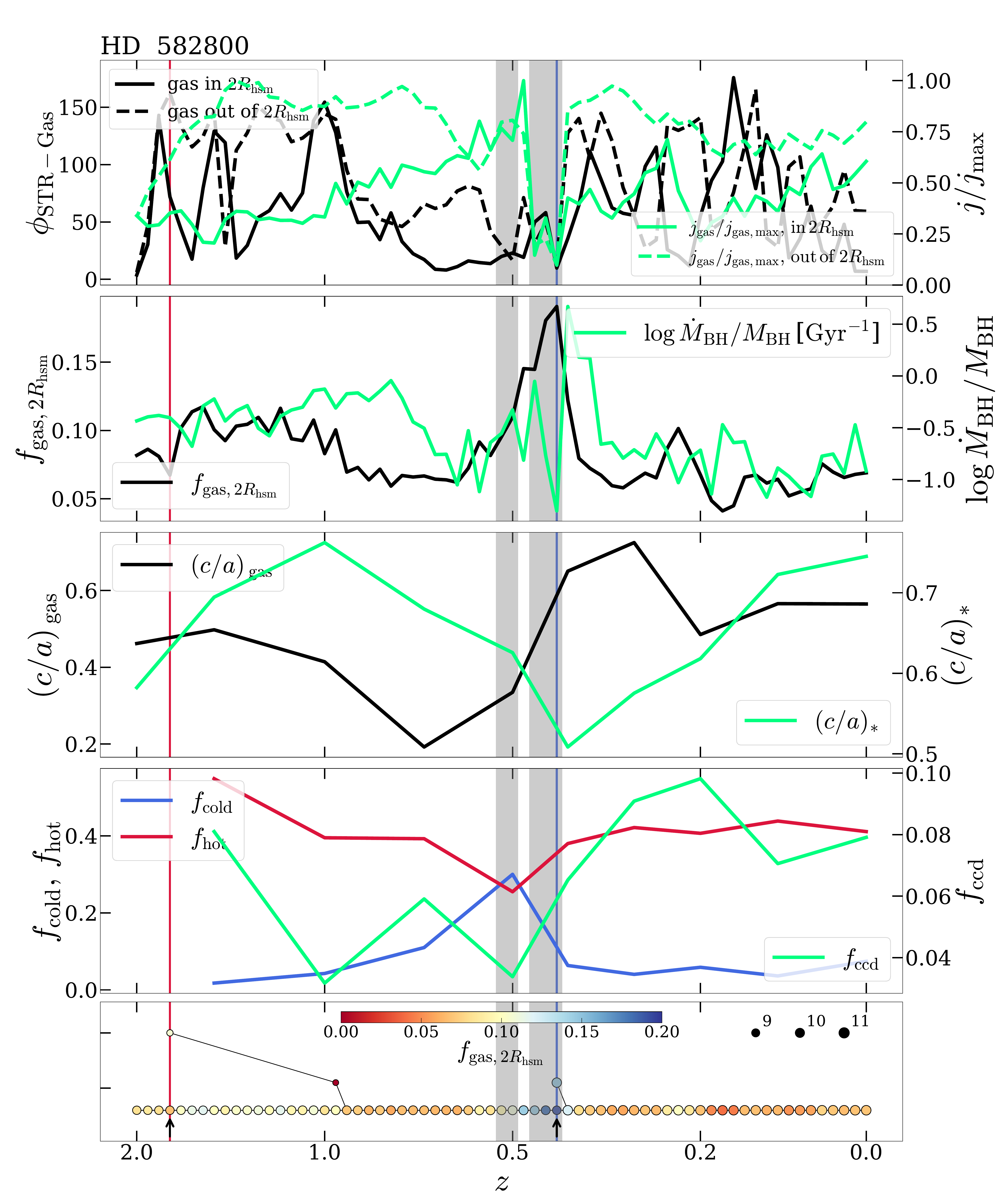}
\includegraphics[width=1\columnwidth]{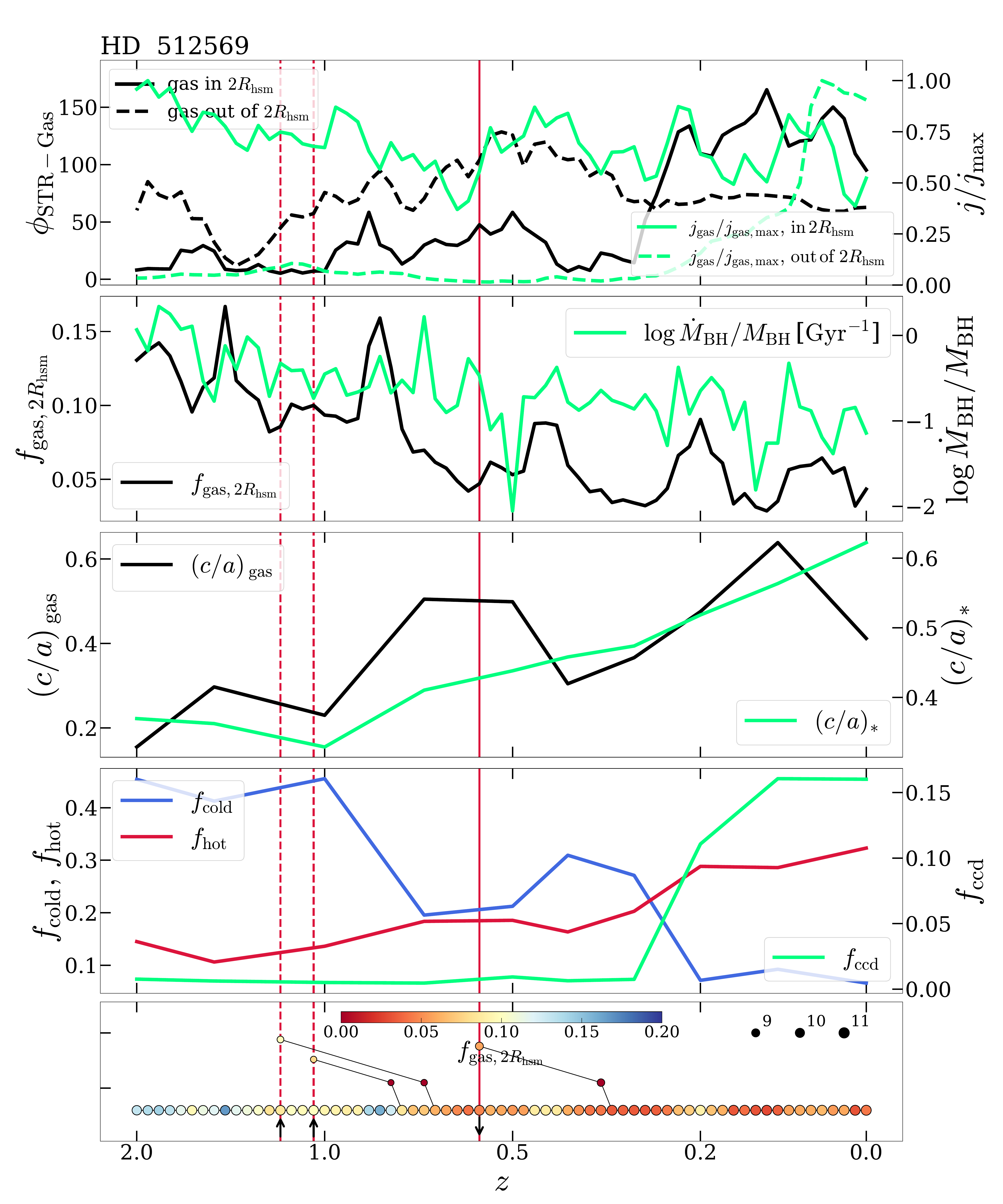}
\caption{The redshift evolution ($0\,\leqslant\,z\,\leqslant\,2$) of
  properties for the two HD examples (IDs: 582800, 512569). The symbols
  are the same as in Fig.~\ref{fig:evol_ND}. The grey shaded regions
  indicate the period when the host galaxy is not a central galaxy.}
\label{fig:evol_HD}
\end{figure*}

Figure~\ref{fig:evol_HD} shows the evolution of two example HDs
selected from our sample. Galaxy HD 582800 has always been dynamically
hot and never been able to grow its disk properly. This can be seen
from the evolution of its hot and cold orbit fractions (4th panel,
left), as well as from the circularity distributions in age bins
(Fig.~\ref{fig:examples_circularity}, upper left panel). It is
interesting to note that around $z=0.5$ this galaxy was a satellite of
a bigger system for a while (indicated by the grey shaded regions),
during and after which both the stellar and gas disks became thicker,
and the cold orbital fraction decreased further while the
counter-rotating disk component also started growing. By $z=0$ this
galaxy remains a dynamically hot system.

Galaxy HD 512569 has experienced a different evolutionary history. It
started with being a well established disk at $z \sim 1$ with a
dominant (low) fraction of cold (hot) stellar orbits and thin
morphologies in both stellar and gaseous disks. Two mergers that
initiated just before $z=1$ and finished by $z=0.7$ have caused a
significant decrease of the cold orbit fraction as well as an increase
of disk thickness, which carry through its following
evolution. Meanwhile, the misalignment between the gas and stellar
spins started building up. Shortly after another merger event which
started at $z\sim 0.5$ and finished at $z\sim 0.3$ with a retrograde
in-coming satellite, the counter-rotating orbit fraction increases (as
can also be seen in the age-binned circular distribution in the bottom
left panel of Fig.~\ref{fig:examples_circularity}), and the
misalignment started growing significantly. By the current epoch, this
system has lost its rotation-supported thin disk and has become
dynamically hot.

The example hot disks show two typical but different evolutionary
paths: in the former case, a normal flat and rotation supported
disk never succeeded to grow, while in the latter case, a normal disk
at high redshift became hot at some point due to a galaxy
interaction/merger. In some cases, a normal disk can first go through
a phase of CR before eventually becoming a HD. In some other cases,
like HD 512569, the outcome that the galaxy has a high fraction of hot
orbits essentially arises as a consequence of a significant growth of
counter-rotating orbits in its most recent history. We stress that
being dynamically hot is mostly a transient state depending on local
conditions of galaxy merger/interactions, rather than an age-dependent
evolution phase.

It is worth noting that the correlations between the emergence of hot
kinematic signatures (morphological -- in 3rd panel, and orbital -- in
4th panel) and strong AGN feedback activities (as indicated by the high
black hole accretion rates; 2nd panel, right) are not evident in most
dynamically hot galaxies. For Galaxy HD 582800 and a few other
galaxies alike, sudden increases in the black hole accretion rate are
always accompanied by merger events. This may indicate that the
activity of central black holes and AGN feedback may not have played a
leading role in the formation of the dynamically hot disk galaxies at
the investigated mass scale.

\subsection{Two example CR galaxies}

\begin{figure*}
\centering
\includegraphics[width=1\columnwidth]{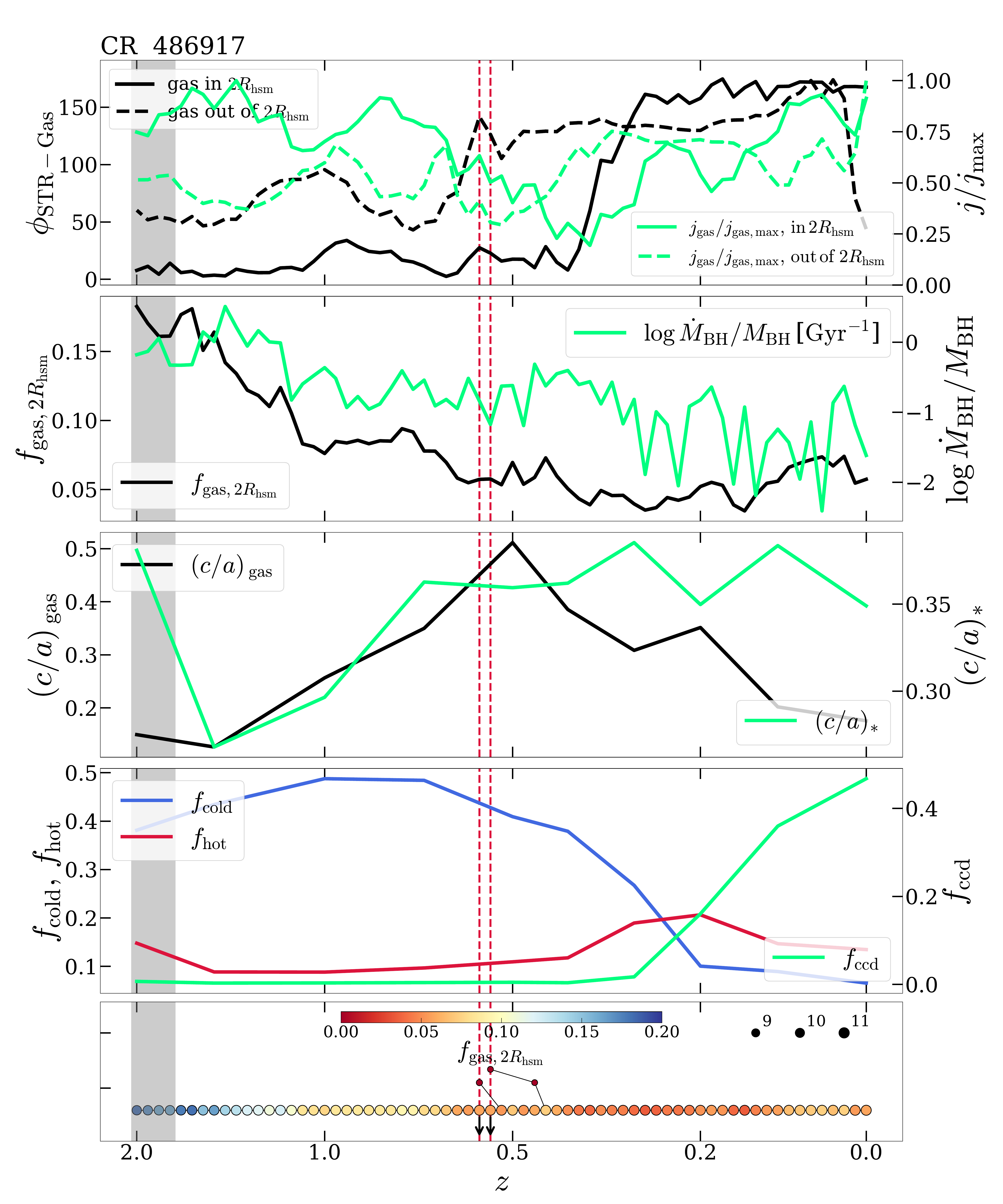}
\includegraphics[width=1\columnwidth]{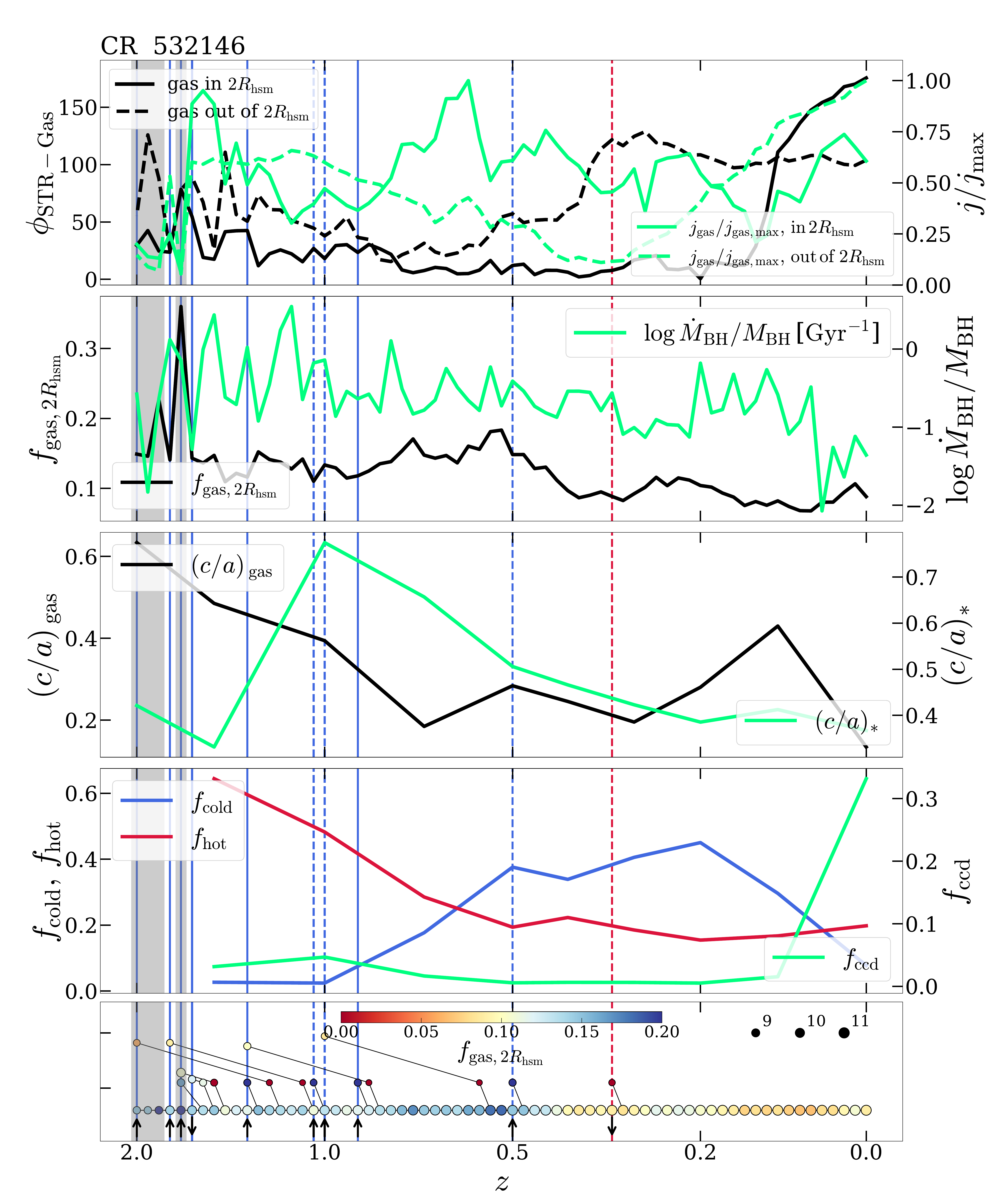}
\caption{The redshift evolution ($0\,\leqslant\,z\,\leqslant\,2$) of
  properties for the two CR examples (IDs: 486917, 532146). The symbols
  are the same as in Fig.~\ref{fig:evol_ND}. The grey shaded regions
  indicate the period when the host galaxy is not a central galaxy.}
\label{fig:evol_CR}
\end{figure*}

Fig.~\ref{fig:evol_CR} shows the evolution of two example CR galaxies
in this study. Galaxy CR 486917 had has a rotation-supported thin
stellar disk for quite some time in its history until just before
$z \sim 0.5$, when two successive retrograde mergers happened, shortly
after which the SFR shot up temporarily, the cold orbital fraction
started decreasing, and misalignment begun to grow
significantly. With another delay in time, the counter-rotating orbit
fraction started increasing by $z \sim 0.2$, and the system finally
became a CR at the current epoch. For Galaxy CR 532146, what
terminated the steady growth of its rotation-supported disk was a
recorded retrograde merger that happened at $z \sim 0.3$, after which
the disk became thicker, and both the gas-star misalignment and the
counter-rotating stellar orbits built up.

It is worth noting that correspondences between retrograde merger
events (shown by down arrows) and the subsequent build up of
counter-rotating orbital fractions are clearly seen in many CR
systems. Very often, a decrease in gas spin and an increase first in
the outer-gas misalignment angle are observed to go together, prior to
and during a large-angle merger event, which is then followed by a
(delayed) rise in the inner-gas misalignment angle as well as a
rebuild of gas spin (after counter-rotating gas has been substantially
accreted to the galaxy center). During this process, the central gas
fraction would decrease after the merger event but the change is not
always sharp. In other words, a clear phase of gas removal, as found
by~\citet{Starkenburg_et_al.(2019)} is not always present in the CR
galaxies in our study. It is also worth noting that, in \citet{Khoperskov_et_al.(2021)}, 25 counter-rotating TNG galaxies of a similar mass range as in this work were studied. They also discovered the connection between the counter-rotation of the stellar component and the external gas infall which captures and mixes with the pre-existing gas, resulting in counter-rotating gaseous disks.

\subsection{Merger statistics}
\label{sec:merger_statistics}
From examining individual cases (as demonstrated for the example
galaxies above), we realize that mergers may have played an important
role in the formation of dynamically hot galaxies and in causing
morphological and kinematic differences among the three galaxy
types. In order to reveal these connections in a statistical fashion,
we further define merger types according to merger orbit angle
$\theta_{\rm orb}$, merger mass ratio $\mu$ and merger gas fraction
$f_{\rm g}$, and carry out merger number counts (per galaxy) since
$z=1$ for the different merger types. We note here that we have also carried out the same analysis on the mergers from $z=2$ to $z=0$ and found that the main results are nearly unchanged, except that the discrepancies between the three types of galaxies become smaller. That is because galaxies of the studied mass range may have experienced similar merger histories at higher redshift (i.e. $z>1$). In addition, the progenitors of the studied galaxies are getting close to the resolution limit at these high redshifts, preventing us from extracting galaxy properties reliably. All analyses hereafter are thus carried out for mergers since $z=1$. The result is presented in Table~\ref{table:table2}.

As can be seen, NDs experienced on average the most frequent galaxy
mergers among all three galaxy samples, CRs experienced the least, and
HDs lie in between. One significant difference among them is the
fraction of merger events where the in-coming galaxy merges with an
orbit angular momentum opposite to that of the main galaxy spin (i.e.,
$\theta_{\rm orb}\,\geqslant\,90^{\circ}$): this fraction is as high
as 57\% among CRs but only 14\% for NDs and 24\% for HDs. It is indeed
seen in over half of the CR cases, as well as some of the HDs, where
gas-star misalignment and growth of counter-rotating orbits emerge
shortly after a retrograde merger event (as is seen for CR 486917, CR
532146 and HD 512569). It is then not surprising to find that the
coherent rotations of NDs have survived the (albeit more frequent)
pro-grade mergers, which make up 86\% of averaged merger events for a ND
galaxy.

The average numbers of mergers which have mass ratio
$\mu\,\geqslant\,0.01$ also differ markedly from nearly half (46\%)
for HDs, 39\% for CRs, to 28\% for NDs. This might serve as
marginal evidence that larger-mass ratio merger events are more likely
to induce the hot dynamical configuration of a galaxy, however, on an
individual bases this connection is not clearly established.

It is interesting to note that CRs have the largest fractions in
gas-rich mergers (i.e., $f_{\rm g}\,\geqslant\,0.1$), up to 70\%,
followed by 60\% for NDs and 51\% for HDs. The supply of new gas
brought in due to mergers is however crucial for newer generations of
star-formation. In the case of CRs, these gas-rich merger events
together with their retrograde merging fashion, may have provided the
basis for the counter-rotating newer-generations of stellar
populations.

\begin{table*}
  \caption{The {\it average} number of mergers since $z=1$ per galaxy
    within different galaxy samples of HDs, CRs, and NDs. See Section~\ref{sec:how_form} for definitions of $\theta_{\rm orb}$, $\mu$, and $f_{\rm g}$.}
\setlength{\tabcolsep}{6.5mm}
\begin{tabular}{cccc}
\hline
\hline
   & HDs & CRs & NDs\\
\hline
\hline
$\theta_{\rm orb}\,\geqslant\,90^{\circ}$ & 0.62 (24\%) & 0.75 (57\%) & 0.50 (14\%)\\
$\theta_{\rm orb} <90^{\circ}$ & 1.97 (76\%) & 0.57 (43\%) & 3.03 (86\%)\\
\hline
$\mu\,\geqslant\,0.01$ & 1.19 (46\%) & 0.51 (39\%) & 1.00 (28\%)\\
$\mu<0.01$ & 1.40 (54\%) & 0.81 (61\%) & 2.53 (72\%)\\
\hline
$f_{\rm g}\,\geqslant\,0.1$ & 1.32 (51\%) & 0.92 (70\%) & 2.11 (60\%)\\
$f_{\rm g}<0.1$ & 1.27 (49\%) & 0.40 (30\%) & 1.42 (40\%)\\
\hline
\hline
Total & 2.59 & 1.32 & 3.53\\
\hline
\hline
\end{tabular}
\vspace{2mm}
\label{table:table2}
\end{table*}

\section{Real-world counterparts}
\label{sec:observation}
In this section, we present a sample of observed late-type galaxies
which have misaligned gaseous and stellar spin axes from the MaNGA
Product Launch-7 (MPL7), which contains 4621 galaxies and is identical
to the MaNGA data included in the SDSS data release 15 (DR15;
\citealt{Aguado_et_al.(2019)}). We show that these gas-star misaligned
galaxies are very plausible real-world counterparts of our HDs and CRs
selected from the IllustrisTNG Simulation according to their stellar
orbit features.

To get the type and misalignment angle of MaNGA galaxies, we follow
the methods of \citet{Jin_et_al.(2016)}, where star-forming galaxies
are identified according to their locations on the
$\log\,\mathrm{SFR}-\log\,M_{\ast}$ plane and the misalignment angles
are defined as
$\Delta\mathrm{PA}=\left|\mathrm{PA}_{\ast}-\mathrm{PA}_{\rm
    gas}\right|$ (where $\mathrm{PA}_{\ast}$ and
$\mathrm{PA}_{\rm gas}$ are the position angles of stars and gas,
measured using the {\sc FIT\_KINEMATIC\_PA} routine described in
\citealt{Krajnovic_et_al.(2006)}). Thus, the misaligned star-forming
galaxies are defined to be galaxies with
$\log\,\mathrm{SFR}>0.86\times \log\,M_{\ast}-9.29$ and
$\Delta \mathrm{PA}>30^{\circ}$ (see \citealt{Jin_et_al.(2016)},
section 2.2.2 and 2.2.3). The velocity field data used here are
extracted from the IFU spectra using the MaNGA data analysis pipeline
(DAP; \citealt{Westfall_et_al.(2019)}) by fitting absorption lines,
making use of the {\sc ppxf} software
\citep{Cappellari_and_Emsellem(2004),Cappellari(2017)} with a subset
of the {\sc MILES}
\citep{Sanchez-Blazquez_et_al.(2006),Falcon-Barroso_et_al.(2011)}
stellar library, {\sc MILES-THIN}. Before fitting, the spectra are
Voronoi binned \citep{Cappellari_and_Copin(2003)} to
$\mathrm{S/N}=10$. Besides, we derive the extended stellar population
properties from the MaNGA Pipe3D Value Added Catalog (VAC;
\citealt{Sanchez_et_al.(2018)}). Redshift and effective radii of these
galaxies are derived from the NASA Sloan Atlas (NSA)
catalog\footnote{\url{http://nsatlas.org/data}}~\citep{Blanton_et_al.(2011)}. After
combining the above catalogs and excluding merging galaxies, we have
55 gas-star misaligned star-forming galaxies with available kinematic
and stellar population properties, with their mass range being
$9.5<\log\,M_{\ast}/\mathrm{M_{\odot}}<10.3$, roughly matching the mass range of our
simulated dynamically hot disk galaxies. To make a comparison, we also
select a sample (with the same galaxy number as the misaligned sample) of
well-aligned star-forming galaxies which have
$\Delta \mathrm{PA}<30^{\circ}$ in the same mass range. The catalog of
these gas-star misaligned galaxies will be released in Chen et al. (in
prep.).

We note that our TNG samples explicitly include flatter galaxies (roughly $L_{\rm dev}<0.5$ and S{\'e}rsic index $n_{\rm S\acute{e}rsic}<2.5$) by design. In comparison, the misaligned MaNGA galaxy sample taken for this study spans a wider range in S{\'e}rsic index, with a median of 2.76, spreading from 1 to 6. It is worth noting that dynamically hot disks acquire thicker morphologies along with hotter stellar kinematics after being perturbed by mergers (as can be seen in Fig~\ref{fig:evol_HD}); a thick disk or even spheroidal shape will be a natural morphology in the later evolutionary stages of these galaxies, which are not investigated directly in this study but are expected to feature the same hot kinematics. We have therefore included all galaxies in this MaNGA sample in our analysis regardless of their S{\'e}rsic index, in order to increase the sample size. We have checked and confirmed that the distribution of the misalignment angle for the adopted MaNGA galaxy sample does not show dependence on the S{\'e}rsic index. To make a direct comparison with MaNGA samples, we project our simulated galaxies (NDs, HDs, and CRs) along the X-axis of the simulation box (equivalent to projecting randomly) and re-calculate the relative parameters (e.g. the $V_{\ast}/\sigma_{\ast}$ profiles and stellar ages in given apertures).

\begin{figure}
\centering
\includegraphics[width=\columnwidth]{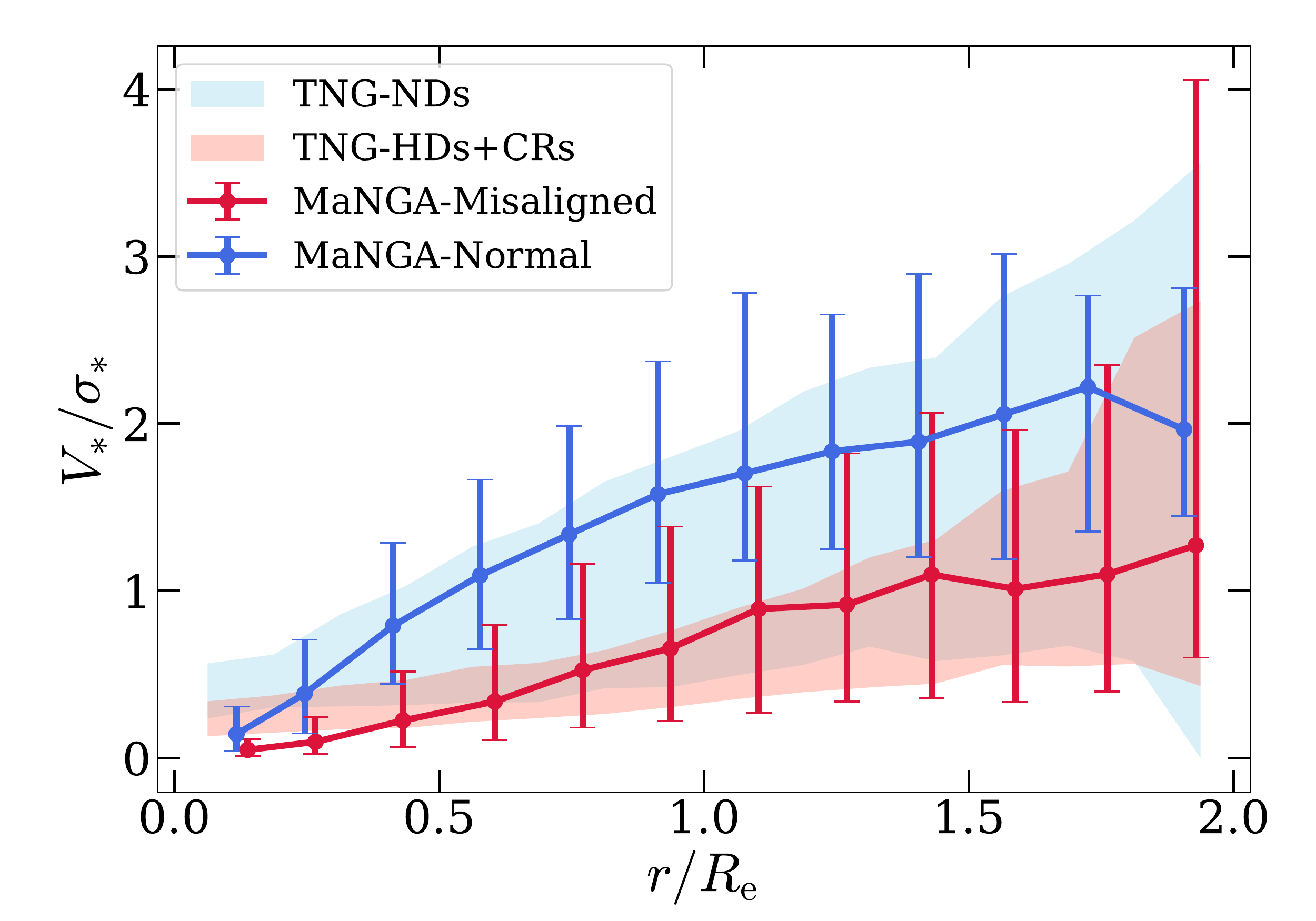}
\caption{The $V_{\ast}/\sigma_{\ast}$ radial profiles ($0-2R_{\rm e}$) of normal (blue) and misaligned (red) samples in MaNGA with the error bars indicating the range from the 16th to the 84th percentiles ($1\sigma$). The shaded regions indicate the 1$\sigma$ ranges of the $V_{\ast}/\sigma_{\ast}$ radial profiles ($0-2R_{\rm e}$) for NDs (blue) and the combination of HDs and CRs (red).}
\label{fig:v2sigma_MaNGA}
\end{figure}

In Fig.~\ref{fig:v2sigma_MaNGA}, we present the $V_{\ast}/\sigma_{\ast}$ profiles of both MaNGA and TNG (X-projection) samples. As can be seen, MaNGA normal and misaligned samples show similar $V_{\ast}/\sigma_{\ast}$ profiles as TNG NDs and HDs+CRs, respectively. Besides, the misaligned star-forming galaxies in MaNGA appear to have lower $V_{\ast}/\sigma_{\ast}$ at any radius than their normal galaxy counterparts, indicating the weaker coherent rotation in misaligned MaNGA galaxies, consistent with the kinematic feature of HDs and CRs in the IllustrisTNG Simulation. We note that due to the random projections for the MaNGA and TNG samples, the $V_{\ast}/\sigma_{\ast}$ is on average lower than that in Fig.~\ref{fig:v2sigma} (evaluated from the edge-on projection) at the same radius.

\begin{figure*}
\centering
\includegraphics[width=2\columnwidth]{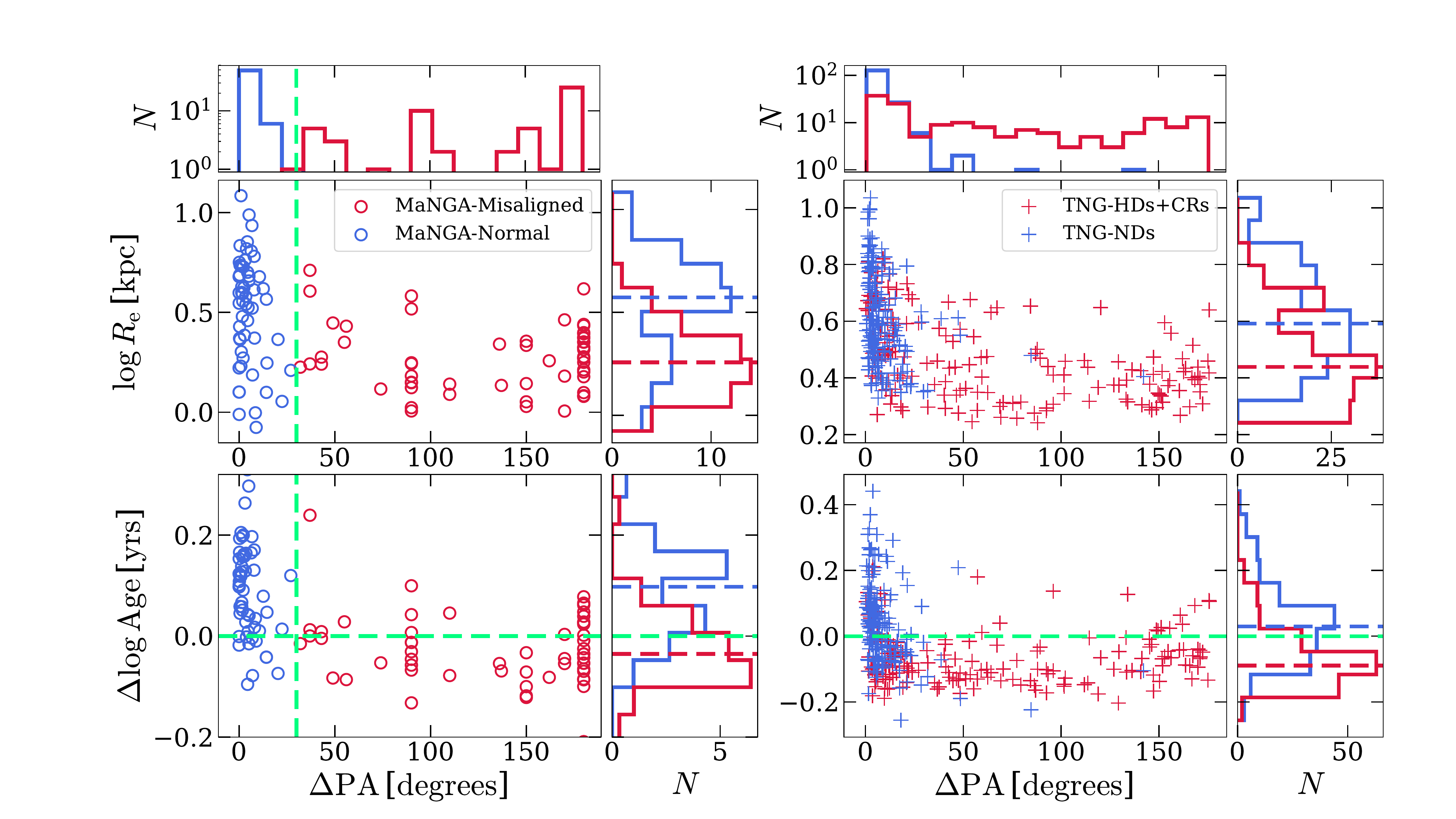}
\caption{Distributions of galaxy size (top panels) and the age difference between two different apertures ($\Delta\log \mathrm{Age}$, bottom panels) as a function of $\Delta \mathrm{PA}$ (see Section~\ref{sec:observation} for definition) of MaNGA (left panels) and TNG (right panels) samples. For MaNGA samples, galaxy size is quantified by the 2D effective radius $R_{\rm e}$ and $\Delta\log\mathrm{Age}$ is calculated within two apertures with radii $0.5R_{\rm e}$ and $R_{\rm e}$; for TNG samples, galaxy size is approximated by the half stellar mass radius ($R_{\rm hsm}$), $\Delta\mathrm{PA}$ is approximated by $\phi_{\rm Gas-STR}$ (see Section~\ref{sec:kin_of_gas_and_star} for definition), and due to the spatial resolution limitation, $\Delta\log\mathrm{Age}$ is calculated within apertures with radii $R_{\rm hsm}$ and $2R_{\rm hsm}$. MaNGA misaligned and normal samples (see Section~\ref{sec:observation} for definitions) are indicated by red and blue circles, respectively; the dynamically hot samples (including HDs and CRs) in TNG are indicated by red plus symbols and NDs are indicated by blue symbols. The green dashed vertical lines in the left panels represent $\Delta \mathrm{PA}=30^{\circ}$, which is the classification criterion of misaligned and normal sample in MaNGA. The green dashed horizontal lines in the bottom panels indicate $\Delta\log\mathrm{Age}=0$, below which galaxies are younger in the center than in the outer parts. The histograms indicate the distributions of $\Delta\mathrm{PA}$ (in the top of the figure) and the investigated properties ($\log\,R_{\rm e}$ and $\Delta \log\mathrm{Age}$, to the right of each panel).}
\label{fig:props_MaNGA}
\end{figure*}

Apart from the similar kinematic feature in MaNGA misaligned star-forming galaxies and TNG dynamically hot star-forming disks (including HDs and CRs), we further demonstrate the connection of these two kinds of galaxies from the other two aspects (i.e. the morphology and the stellar population). We present, in the left panels of Fig.~\ref{fig:props_MaNGA}, the distributions of galaxy size and $\Delta\log\mathrm{Age}$ as a function of $\Delta\mathrm{PA}$ for MaNGA misaligned (red) and normal (blue) galaxies. For these MaNGA samples, galaxy size is quantified by the 2D effective radius $R_{\rm e}$ and $\Delta\log\mathrm{Age}$ is the age difference between two apertures with radii $0.5R_{\rm e}$ and $R_{\rm e}$ (negative $\Delta\log\mathrm{Age}$ means that stellar populations in the galaxy center are younger than at the outskirts). As can be seen, MaNGA samples show decreasing trends for $\log\,R_{\rm e}$ and $\Delta\log\mathrm{Age}$ towards high $\Delta\mathrm{PA}$. More specifically, the misaligned star-forming galaxies ($\Delta\mathrm{PA}>30^{\circ}$) in MaNGA are obviously of smaller sizes, and have younger stellar populations in their central regions ($\Delta\log\mathrm{Age}<0$), reflecting recent star-formation happening therein. These results for MaNGA galaxies are all consistent with the results for the TNG samples (HDs+CRs vs. NDs), which are shown in the right panels. We note here that for the TNG samples, galaxy size is approximated by the half stellar mass radius, $R_{\rm hsm}$, $\Delta\mathrm{PA}$ is approximated by $\phi_{\rm Gas-STR}$ (see Section~\ref{sec:kin_of_gas_and_star} for definition), and due to the spatial resolution limitation, $\Delta\log\mathrm{Age}$ is calculated within apertures with radii $R_{\rm hsm}$ and $2R_{\rm hsm}$. Besides, HDs, CRs, and NDs in the IllustrisTNG Simulation are divided by orbital fractions (rather than $\Delta\mathrm{PA}=30^{\circ}$ used in MaNGA), and as a result, the $\Delta\mathrm{PA}$ distributions of TNG dynamically hot disks and their normal counterparts are not so well separated as the MaNGA samples.

Despite the slight difference of parameters used in MaNGA and TNG, we can still confirm that the misaligned star-forming galaxies in MaNGA show features in kinematics, morphology, and stellar populations that are all consistent with the dynamically hot star-forming disk galaxies (HDs and CRs) in the IllustrisTNG Simulation, strongly suggesting that these misaligned star-forming MaNGA galaxies are likely to be the real-world counterparts of HDs and CRs in IllustrisTNG. We also predict that the HI disk morphologies of these gas-star misaligned galaxies would be thicker, more concentrated and less well established (as shown by Fig.~\ref{fig:gas_mor}), and that the newly formed stars co-rotate with the existing (counter-rotating) gaseous disk. In particular, for large-angle gas-star misaligned galaxies (that correspond our counter-rotating galaxies), the spin directions of their youngest stellar populations are expected to line up with the cold gas, with both being opposite to that of the older populations in these systems. This should also be reflected in the kinematic maps of H$_{\alpha}$, HI, and stars.

\section{Conclusions and Discussion}
\label{sec:conclusion}

In this study, we investigate the nature and origin of (lower-mass)
dynamically hot but still actively star-forming disk galaxies in the
IllustrisTNG Simulation. These galaxies are selected according to
their star-formation activities, stellar morphologies as well as stellar
orbital fractions (see Section~\ref{sec:selection}). In particular, we
identify two subsets of these dynamically hot systems, i.e., hot disk
(HD) galaxies and counter-rotating (CR) galaxies (see
Fig.~\ref{fig:ltgs} and Fig.~\ref{fig:forbits}). Our main conclusions
are:
\begin{enumerate}
\item Dynamically hot star-forming disk galaxies show smaller sizes
  (and as a result, higher surface stellar mass densities), thicker
  morphologies (higher $(c/a)_{\ast}$, $(c/a)_{\,\rm gas}$, and
  $L_{\rm dev}/L_{\rm tot}$, see Fig.~\ref{fig:examples_morphology},
  \ref{fig:stellar_mor}, and \ref{fig:gas_mor}), more
  centrally-concentrated on-going star-formation (see
  Fig.~\ref{fig:examples_SFR}) and as a result, larger age gradients
  towards younger stellar populations in the center than at the
  outskirts (see Fig.~\ref{fig:sp_profiles}), in comparison to their
  normal disk counterparts.
    
\item Dynamically hot star-forming disk galaxies show a common feature
  of hosting kinematically misaligned gas and stellar disks (see
  Fig.~\ref{fig:examples_kinematics} and
  Fig.~\ref{fig:spin_misalignment}), which show weaker coherent
  rotations and markedly smaller spins than their normal star-forming
  disk counterparts (see Fig.~\ref{fig:spins}, \ref{fig:v2sigma}, and
  \ref{fig:examples_circularity}).
    
\item Galactic interactions and mergers may have played a leading role
  in the formation of the dynamically hot configurations (see
  Table~\ref{table:table2} and Section~\ref{sec:merger_statistics} for
  details): CRs experienced the least frequent mergers -- on average
  1.32 mergers per galaxy between $0\,\leqslant\,z\,\leqslant\,1$,
  among which 57\% (w.r.t 14\% for NDs and 24\% for HDs) showed
  retrograde merger orbits and 70\% (w.r.t. 60\% for NDs and 51\% for
  HDs) are gas-rich events; HDs experienced on average 2.59 mergers
  per galaxy (within the same redshift range), and nearly half of them
  (46\%, w.r.t. 39\% for CRs and 28\% for NDs) are high mass-ratio
  events. In comparison, their normal disk counterparts experienced
  the most frequent mergers -- 3.52 mergers per galaxy, however retaining disk morphology and cold kinematics. We emphasize that retrograde mergers as well as larger-mass ratio mergers are primarily responsible to the subsequent build-up of disk thickness and hot/counter-rotating orbits in these galaxies (see Figs.~\ref{fig:evol_HD} and \ref{fig:evol_CR}). This is similar to the situation at the more massive end for early-type galaxies for which such mergers play a key role in the formation of slow-rotators during binary mergers (\citealt{Bois_et_al.(2011)}; similar studies also see \citealt{Martig_et_al.(2009)} and \citealt{Hoffman_et_al.(2010)}). Observational connections between counter-rotating galaxies and retrograde mergers have long been found, such as for NGC 3593, NGC4550 and NGC4191~\citep{Coccato_et_al.(2013),Coccato_et_al.(2015)}. In contrast, a
  correspondence between active AGN feedback and the emergence of
  dynamically hot disks in these galaxies is not readily evident (also see \citealt{Khoperskov_et_al.(2021)}). We
  stress that being dynamically hot appears largely as a transient
  state related to mergers and other galaxy interactions, rather than
  arising as an age-dependent evolutionary phase of star-forming disk
  galaxies.
    
\item Gas-star misaligned star-forming galaxies from the MaNGA observations show features in morphology, kinematics and stellar populations that are consistent with the dynamically hot star-forming disks from the TNG simulation. In particular, they are also found to have weaker (co-)rotation and younger stellar populations in galaxy centers than at the outskirts (see Fig.~\ref{fig:v2sigma_MaNGA} and Fig.~\ref{fig:props_MaNGA}), indicating that these gas-star misaligned star-forming galaxies are the real-world counterparts of the dynamically hot TNG disk galaxies examined in this study. In addition, we also predict that the HI disk morphologies in these systems should be thicker, more concentrated and less well established, and that the newly formed stars co-rotate with the existing (counter-rotating) gaseous disk. It is worth noting that in the study of \citet{Khoperskov_et_al.(2021)}, 90\% of counter-rotating stars were found to form in-situ from counter-rotating gaseous disks. We propose for large-angle gas-star misaligned galaxies (that correspond our counter-rotating galaxies), the kinematic maps of H$_{\alpha}$, HI, and stars should reveal coherent rotations of the youngest stellar populations and the cold gas, both of which are spinning in the opposite direction with respect to the bulk stellar populations in these systems.
\end{enumerate}

\subsection{Minorities in a bigger picture}

Star-forming and rotation-supported disk galaxies manage to maintain
higher spin than their dynamically hot counterparts.  The former
dominate in number among today's actively star-forming galaxies, while
the latter -- systems that possess a good portion of non-corotating
stellar components, such as the hot disks and counter-rotators in this
study -- are only minorities. This is not something that should be
taken for granted. It is interesting to put this fact into a larger
context and contemplate how this could happen given different
processes that determine a galaxy's spin today.

For a given galaxy/halo, the initial tidal field within its Lagrangian
region serves as a cosmic spindle machine built upon the filamentary
skeletons connected at joints, where flows of gas and dark matter, as
well as galaxies/halos eventually meet. On the one hand, such a field
provides the galaxy, during the linear growth era, with its initial
angular momentum originating from large scale structure torques,
which, during the subsequent gravitational collapse and star
formation, spins up the gas content and builds up a rotational stellar
disk.
 
On the other hand, the large-scale tidal field to a certain degree
also maps out some initial passage ways to numerous galaxy merger
events, in particular major mergers, which transfer their coherent
orbital angular momenta to the galaxy/halo and further spin it up
during such events~\citep{Vitvitska_et_al.(2002), Deason_et_al.(2011),
  Rodriguez-Gomez_et_al.(2017)}. Note that the Lagrangian region (at
the halo's turnaround) in fact already includes most of the structures
that would later-on merge into the
system~\citep{Liao_et_al.(2018)}. In this sense, the initial and
large-scale tidal field has pre-wired the co-rotating and well-aligned
merger/accretion events, and thus predetermines a dominant population
of rotational disk galaxies in the first place (see also
\citealt{Rodriguez-Gomez_et_al.(2017)}, fig. 10).

Mergers that happen later on, most frequently being minor/mini
mergers, and in particular those that involve more massive galaxies,
can then be significantly affected by local and internal torques. On
top, there are statistical effects from multi-body interactions that
induce randomness. As a consequence, these merger trajectories
gradually lose memory of the larger-scale tidal field.  In-coming
structures, either in form of surviving subhalos and satellite
galaxies, or being tidally stripped and smoothly accreted to the
galaxy, can significantly alter the host system spin with their
in-coming orbit angular momenta (like the CRs and HDs in this study),
as well as contribute to the internal torque and induce gravitational
instabilities, affecting the subsequent dynamical evolution of the
host system. The remnants can also serve as a low-angular momentum
reservoir of accreted ex-situ stars. All of the above could reduce the
overall angular momentum budget, and induce hot configurations in
stellar dynamics~\citep{Vitvitska_et_al.(2002),
  Zjupa_and_Springel(2017), Rodriguez-Gomez_et_al.(2016),
  Rodriguez-Gomez_et_al.(2017)}.

Apart from galaxy mergers and merging remnants, which all serve as an
external source for changing a galaxy's spin, the internal structures
and processes of a galaxy can also have an effect. Gravitational
instabilities induced by bars or other high-order density moments make
stars and gas lose their angular momenta and sink to the centers of
galaxies. Energy and momentum feedback from stellar evolution or
super-massive black hole growth, which do not directly affect existing
stellar populations, can however change/destroy spin carried by cold
gas flowing into the galaxy, leaving imprints on the kinematics of the
subsequent generation of stars. Such effects can be more important for
higher-mass galaxies, which are investigated in a parallel work to
this study (Lu et al., in prep).

So what is the answer to the question why there cannot be as many
slow- or non-rotating disks as there are fast-rotating disks {\it
  among actively star-forming} galaxies in the Universe? Because once
the former type comes into existence -- through the above-mentioned
catastrophic processes of spin-reduction -- not only existing stellar
populations but also gas would then quickly (within a dynamical time
of $\ll1$ Gyr) sink to the centers of galaxies, where star-forming
activities shoot up, accompanied by super-massive black hole accretion
and feedback, both diminishing the cold gas reservoir, whose original
angular momentum and coherent motion of accretion are then largely
wiped out due to direct consumption, feedback heating and turbulent
dissipation. Such systems now have heavily reduced star-formation, if
not being completely quenched. This is why active star-formation and a
high-spin rotational disk always come hand in hand.

\subsection{Transition between phases}

It is worth noting that even for a ``healthy'' rotation supported
star-forming disk, there could have been transitional stages during
which the cold gas reservoir was nearly used up and where star
formation died down temporarily, which is indeed seen in some examples
of this study. But not before long, the polluted and ejected hot gas
(as a consequence of feedback) from the inner galaxy meets pristine
cold gas from the CGM/IGM in the outskirts of the galaxy, once again
cooling down and falling back to the inner region, where a new cycle
of star-formation starts up again. In most cases, this new round of
gas accretion ends up co-rotating with the existing stellar disk, both
carrying imprints of the same tidal torque field present on large
scales. With time, a healthy rotational star-forming disk thus manages
its maintenance.

However, there are also cases where the new gas acquisition is
strongly affected by, e.g., galaxy mergers (mainly investigated in
this study), internal torques from stellar bars, disturbances from
energetic feedback, and dissipation when colliding with opposite
streams of gas. All these can significantly reduce the total gas spin
and cause kinematic misalignment between gas and stars. Then we are
back to the picture described above, followed by central star
formation and more energy feedback -- a dynamically hot star-forming
disk emerges!

It is interesting to note that the ensuing fate of these dynamically
hot star-forming disks is in fact variable, in particular, if the
newly accreted gas eventually dominates the total gas motion and
happens to spin in the opposite direction to the existing stellar
disk. As such a system is gravitationally stable, with time more and
more counter-rotating stars born in this gaseous environment would
accumulate and eventually dominate the total stellar spin of the
galaxy. With the total spin flipping its direction (now dominated by
the counter-rotating component prior to this moment), what used to be
the majority stream of co-rotating stars becomes the counter-rotating
minority.

In another extreme case, if the triggered feedback process becomes too
vicious, so much so that the gas remains hot over some long cooling
time, then star-formation will cease. This is why we also observe a
large number of slow- or non-rotating ellipticals with quenched star
formation in the Universe today. In a parallel work to this study, we
investigate such systems at a higher mass scale (Lu et al. in
prep). Notice that in more massive galaxies (in particular for
$M_* > 10^{11}\,\mathrm{M_{\odot}}$), this process is also accompanied
by more frequent minor mergers, which are found to play a dominant
role in diminishing a galaxy's overall spin and aggravating stellar
contraction~\citep{Vitvitska_et_al.(2002),
  Rodriguez-Gomez_et_al.(2017)}. What will happen to these systems? If
we would wait long enough, we might see them turn again into shining
star-forming disks, just like the rejuvenated disk galaxies at high
redshifts, but the present-day slow-down of cosmic structure formation
due to dark energy might also prevent this from happening again.

\section*{Acknowledgements}
We acknowledge Drs. Lan Wang, Yougang Wang, and Shihong Liao for useful comments and Min Bao for providing the misalignment angle data of the MaNGA galaxies. We also thank the referee for constructive and insightful suggestions and comments which improved the paper. This work is partly supported by the National Key Research and Development Program of China (No. 2018YFA0404501 to SM), by the National Science Foundation of China (Grant No. 11821303, 11761131004 and 11761141012). JW acknowledges the support from the National Science Foundation of China (Grant No. 11991052). VS acknowledges partial support from the German Science Foundation, DFG (Grant No. 653578).

\section*{Data availability}
General properties of the galaxies in the IllustrisTNG Simulation is available from \url{http://www.tng-project.org/data/}. The rest of the data underlying the article will be shared on reasonable request to the corresponding author.
%%%%%%%%%%%%%%%%%%%%%%%%%%%%%%%%%%%%%%%%%%%%%%%%%%

\bibliographystyle{mnras}
\bibliography{ref} 

%%%%%%%%%%%%%%%%% APPENDICES %%%%%%%%%%%%%%%%%%%%%

%%%%%%%%%%%%%%%%%%%%%%%%%%%%%%%%%%%%%%%%%%%%%%%%%%

\label{lastpage}
\end{document}